\documentclass[runningheads]{llncs}
\usepackage{graphicx}
\usepackage{cutypes}

\begin{document}
\title{Type-directed Bounding of Collections in Reactive Programs}
\titlerunning{Type-directed Bounding of Collections}
\author{Tianhan Lu, Pavol \v{C}ern\'{y}, Bor-Yuh Evan Chang, Ashutosh Trivedi}
\authorrunning{Lu et al.}
\institute{University of Colorado Boulder\\
%
\email{\{tianhan.lu,pavol.cerny,bec,ashutosh.trivedi\}@colorado.edu}}
\maketitle              
\begin{abstract}
  Our aim is to statically verify that in a given reactive
program, the length of collection variables does not grow beyond a given bound. 
We propose a scalable type-based technique that checks that each
collection variable has a given refinement type that specifies constraints about
its length. 
A novel feature of our refinement types is that the refinements
can refer to {\em AST counters} that track how many times an AST node has been
executed. 
This feature enables type refinements to track limited flow-sensitive
information.
We generate verification conditions that ensure that the
AST counters are used consistently, and that the types imply the given bound.
The verification conditions are discharged by an off-the-shelf SMT solver.
Experimental results demonstrate that our technique is scalable, and effective
at verifying reactive programs with respect to requirements on length of
collections.

  \end{abstract}

\section{Introduction}
\label{sec:introduction}

Collections are widely used abstract data types in programs. Collections, by
providing a layer of abstraction, allow a programmer to flexibly choose
different implementations leading to better modularity essential for developing
good quality software. Since collections are extensively used, related
performance issues have attracted considerable
attention~\cite{olivo2015static,xu2008precise,xu2010detecting}. Besides
performance issues, improper usage of collections may lead to security
vulnerabilities such as denial-of-service (DoS) attacks. The performance and
security issues are more pronounced in reactive programs such as
service threads in operating systems or web applications.
An important category of DoS vulnerabilities is out-of-memory error caused by
collections with excessively large lengths.

\noindent {\bf Problem.}
The goal of this paper is to verify bounds on collection
lengths using a scalable type-directed approach. Given constraints on inputs,
our technique statically verifies at any point of execution total length of
collection variables is less than a given bound.  Verifying bound on collection
lengths for reactive programs brings the following {\em challenges}:
\begin{description}
\item[Non-termination.] Reactive programs do not terminate.  The most common
  method for resource bound analysis is based on finding loop bounds
  ~\cite{carbonneaux2015compositional,giesl2014proving,gulwani2009control,gulwani2010reachability,sinn2014simple,zuleger2011bound}.
  This method therefore does not directly apply to reactive programs.
\item[Scalability.] We need a scalable and modular solution, because real world reactive
  programs such as web servers are large (e.g. up to $30 kloc$).
\item[Non-inductiveness of invariants.] The necessary safety invariants might
  be {\em non-inductive}. For instance, collection lengths of a program may be
  bounded, but this is at first glance not provable by checking each statement
  in isolation, because a particular statement might simply add an element to a collection,
  thus breaking an invariant that is naively constructed to help verifying boundedness.
\end{description}

\noindent {\bf Approach.}
We now describe our approach, with a focus on how the three challenges are
addressed. We develop a refinement type system, where  the user is able to
specify bounds on collection lengths, as well as an overall guarantee on the total
length of all collections. These bounds might be symbolic, referring to for
instance to bounds on lengths of input collections. Our tool \toolname\;then type
checks the program, and proves (or refutes) the overall guarantee.

First, to address the challenges of non-termination, our system relies purely on
{\em safety properties}, never requiring a liveness property such as
termination. We also do not require finding loop bounds.

Second, to address the challenge of scalability, we use type-based reasoning
only. This entails checking at most one invariant per collection, as opposed to
one invariant per each code location (as the approaches based on abstract
interpretation~\cite{gulwani2009control,gulwani2010reachability} might need).

Third, to address the challenge of non-inductiveness of invariants, we allow the
refinement refer to {\em AST counters} that count how many times an Abstract Syntax Tree (AST) node has
been executed. For instance, consider the fragment:
\begin{verbatim}
  while (true) { if (*) { C: s.add(r1);...;D: t.add(r2); }  }
\end{verbatim}
and suppose we are interested  in the invariant $|\len{s}-\len{t}| \le 1$, that is, the difference between lengths of
the two collections {\tt s} and {\tt t} is at most 1. The invariant is not
inductive, the statement {\tt s.add(r)} breaks it. However, let {\tt C} be a
counter associated with the AST node of {\tt s.add(r1)}, and {\tt D} with {\tt
t.add(r2)}. The invariant $\len{s}+D=\len{t}+C$ holds.
We can then add a counter axiom $(D+1 \equiv C)\lor(C \equiv D)$ as the two statements are inside a same basic block. Counter axioms are the place where the limited amount of flow-sensitive information that our system uses is captured. The inductive invariant and the axiom together imply the property we are interested in:  $\mid\len{s}-\len{t}\mid\le1$.

\noindent {\bf Contributions.} The main contributions of this paper are
\begin{itemize}
	\item \textbf{Refinement types for collection lengths.} We propose to encode the
	total length of collection variables as safety properties of all reachable
	program states, as opposed to relying on analyzing time bounds. We develop a
	refinement type system where the refinements allow reasoning about collection
	lengths.
	\item \textbf{AST counters for inductive invariants.} A novel feature of our
	refinement types is that the refinements can refer to {\em AST counters} that
	track how many times an AST node has been executed. This feature enables type
	refinements to track limited flow-sensitive information.
	\item \textbf{Empirical evaluation.} Experimental results show that our
	approach scales to programs up to $30kloc$ ($180kloc$ in total), within 52 second of analysis time per benchmark.
	Moreover, we discovered a Denial-of-Service vulnerability in one of our benchmarks because of correctly not being able to verify boundedness.
\end{itemize}

\vspace{-1em}
\section{Overview}
\label{sec:overview}
We demonstrate our approach for verifying the total collection lengths for
reactive programs on a motivating example in
Figure~\ref{fig:motivating-example}.

\subsection{Using \toolname}
Overall, a user interacts with our tool \toolname\;as follows. First, they  write a driver that
encodes a particular usage pattern that they are interested in.  Then they
specify invariants as type annotations. After these two steps, our type system
will take care of the rest by automatically checking if the invariant relations
are valid. If the invariants are indeed valid, \toolname\;will automatically discharge
a query to an off-the-shelf SMT solver, returning the result ``verified'' or ``not
verified''. The ``verified'' answer is conclusive, as our method is sound.  The
``not verified'' is inconclusive: either the bound does not hold, or the user has
not provided sufficient invariants to answer the verification problem.

\begin{figure}[t]
\begin{lstlisting}[language=java]
void driver(@Inv("true") List<String> input) {
	@Guarantee("len(blogDB)<2")
	@Inv("len(blogDB)=c8-c10") List<String> blogDB = new List<String>();
	@Inv("iterOf(input)") Iterator<String> it = input.iterator();
	String blog;
	while(*) {
		blog = it.next();
		c8:     postNewBlog(blog, blogDB);
		c9:     showBlog(blogDB);
		c10:    deleteBlog(blogDB);
} }
@Summary{"len(blogDB')=len(blogDB)+1"}
void postNewBlog(String blog, List<String> blogDB) {//callback: add post
	blogDB.add(blog);
}
@Summary{"len(blogDB')=len(blogDB)-1"}
void deleteBlog(List<String> blogDB) { //callback: delete last post
	blogDB.remove();
}
@Summary{"len(blogDB')=len(blogDB)"}
void showBlogs(@Inv("true") List<String> blogDB) {
	@Guarantee("len(toShow)<=len(blogDB)+2")
	//callback: display blog contents
	@Inv("len(toShow)-idx(it)=c28+c30+c33-c32") List<String> toShow = new List<String>();
	@Inv("iterOf(blogDB)")Iterator<String> it = blogDB.iterator();
	String blog;
	blog = "Welcome!\n";
	c28:  toShow.add(b);
	blog = "Blog begins:\n";
	c30:  toShow.add(b);
	while(*) {
        c32:    blog = it.next();
		c33:    toShow.add(blog);
	}
// render toShow as an HTML page
}
\end{lstlisting}
\caption{Motivating example: a simplified version of a blogging server.}
\label{fig:motivating-example}
\vspace{-2em}
\end{figure}
\noindent {\bf Example (Blogging server).} We simplified code from a Java web server based on the Spring framework that
allows users to upload a blog post, delete a blog post and render a list of posts as an
html page. Callback methods \code{postNewBlog}, \code{deleteBlog}, and
\code{showBlogs} implement these functionalities. Method
\code{driver} encodes an infinite input sequence that a user of our tool is interested in: it first
reads a blog from input and appends it to the database, then renders the blog as
an HTML page, and finally removes the blog from database. Our goal is to verify
the boundedness of total collection lengths in every method separately, when
input variables satisfy given constraints (e.g., inputs can have upper bounds on their length). In particular,
callback methods \code{postNewBlog} and \code{deleteBlog} do not declare
collection-typed variables and therefore they are vacuously bounded. More
interestingly, we would like to verify the following bounding predicates denoted by \code{@Guarantee} in Figure \ref{fig:motivating-example}
\begin{itemize}
	\item The total length of
collection variables in method \code{driver} is less than 2, i.e.
$\len{blogDB}<2$
  \item Total length of collection variables in method
\code{showBlogs} is less than or equal to length of input variable
\code{blogDB}, i.e. $\len{toShow}\le\len{blogDB}+2$
\end{itemize}
We emphasize that our approach is able to verify above bounds when there exist
neither time bounds nor input bounds, because input variables \code{input} and
\code{blogDB} have no constraint at all, i.e. a $\kwtrue$ constraint.

The notation \code{@Inv} in Figure \ref{fig:motivating-example} denotes a
refinement type. The content inside the brackets following \code{@Inv} is the
refinement of that particular type. For example, $\len{blogDB}=\cnt{\dbadd}-\cnt{\dbrmv}$ is a type
refinement on variable \code{blogDB}.

\noindent {\bf Specifying invariants with AST counters.}
We now explain the role of the AST counters in the invariant. For example, if we
look at the inner loop at \exinnerwhile\;in Figure \ref{fig:motivating-example},
the property we most likely need for list \code{toShow} is
$\len{toShow}\le\iter{it}+2$, where $\iter{it}$ represents the number of
elements that has been visited using iterator $it$. However, this property is
actually not inductive because it breaks after line \showadd\;(as well as line \showaddtwo), as
$\len{toShow}$ is incremented by 1 but nothing else is updated in the invariant.
However, we can add AST counters to the invariant, and obtain
$\len{toShow}-\iter{it}=\cnt{\showadd}+\cnt{\showaddtwo}+\cnt{\showwhileadd}-\cnt{\itnext}$. We thus obtain an inductive invariant that
is then used as the type of {\tt toShow}.

The purpose of these counters is to enable writing expressive invariants. The
interesting invariants usually do not depend on the value of the counters (the
value grows without bound for nonterminating programs), just on relations
between counters of different AST nodes. These could be seen on the example in
the previous section.

As another example, consider how we reason about the non-terminating loop at
\exwhile, we first summarize the effects of callback \code{postNewBlog} and
\code{deleteBlog} on any collection variable passed in as argument(s), which is
to add 1 element to or remove 1 element from list \code{blogDB}. Method
summaries are automatically applied at invocation sites. Next, since we have AST
counters, we are now able to easily define the length of variable \code{blogDB}
as an inductive invariant $\len{blogDB}=\cnt{\dbadd}-\cnt{\dbrmv}$ (shown at
line \invdb) that hold at before and after every execution step. Note that this
invariant serves as a safety property of all program states under the existence
of non-terminating executing traces, which is the root cause of the mainstream approach in resource bound analysis to fail under the scenario of reactive
programs.

\subsection{Inside \toolname}

\noindent {\bf Typechecking.} Our type system is based on Liquid types \cite{rondon2008liquid}, where the
refinements can express facts about collections and AST counters. Our type
checking rules are standard, with added rules that capture the semantics of
collections (lists) and counters.

\noindent {\bf Constraints on AST counters.} Constraints on AST counters are
generated from the Abstract Syntax Tree structure of the program. For instance,
AST counter \code{\cnt{\itnext}} is always either greater than (by $1$)
AST counter \code{\cnt{\showwhileadd}} (after executing line \itnext) or equal
to it (after executing line \showwhileadd) at any time during an execution. We
formalize this and other relations on counters in a set of axioms.

\noindent {\bf Verification condition generation.}
We generate verification conditions that ensure that the AST counters are used
consistently, and that the types imply the given bound. For instance, now that
we have invariants describing lengths of list \code{blogDB} and \code{toShow} in
the method \code{showBlogs}, we can plug in counter axioms and check the
required implications. For instance, the type of \code{toShow} is
$\len{toShow}-\iter{it}=\cnt{\showadd}+\cnt{\showaddtwo}+\cnt{\showwhileadd}-\cnt{\itnext}$. From the counter axioms, we have
that $\cnt{\showadd}\le1\land\cnt{\showaddtwo}\le1$ (as the corresponding statements are executed once at most) and
$(\cnt{\itnext}\equiv\cnt{\showwhileadd}+1)\lor(\cnt{\itnext}\equiv\cnt{\showwhileadd})$ (as the corresponding statements are sequentially executed). We then use an off-the-shelf SMT solver to check that the inductive
invariant and the counter axioms imply the guarantee that the user specified:
$\len{toShow}\le \len{blogDB}+2$.

\vspace{-1em}

\section{\toolname\;Type System}
\label{sec:definition}
In this section, we present the core calculus of our target program along with
the types and refinements, and operational semantics.
As usual, we write $\mathbb{B}$ and $\mathbb{Z}$ for the Boolean and integer domains. 
We write $\overline{v}$ to denote a list of syntactic elements separated either by comma or semicolon: $v_1, v_2, \ldots, v_k$ or $v_1; v_2; \ldots; v_k$.
We also write $\listappend{\overline{v}}{v_{k+1}}$ for the list value $(v_1, v_2, \ldots, v_k,
v_{k+1})$.
We model other types of collection data types (such as sets and maps) as lists because of being only interested in sizes of collection-typed variables.

\subsection{Syntax and Refinement Types}
\label{subsec:program-syntax}
\label{subsec:type-system}
\begin{figure}[t]
  \centering
  \subfloat[The core calculus.]
  {
           \label{fig:syntax}
  \begin{tabular}{ l L L L }
    Method definition & M & \bnfdef & 
    \overline{\edecl{\ty}{u}}\;
    \overline{\edecl{\ty}{x} = e}\;
    \stmt \\
    Compound statements & \stmt & \bnfdef &
    \stmt_\text{B}
    \bnfalt \eblock{\stmt}
    \bnfalt \eif{\expr}{\stmt_1}{\stmt_2}
    \bnfalt \ewhile{\expr}{\stmt}
    \\
    Basic statements & \stmt_\text{B} & \bnfdef &
    x = \expr
    \bnfalt x = \enext{z}
    \bnfalt \eremove{y}
    \bnfalt \eadd{y}{x}
    \bnfalt \eskip
    \\
    Expressions & \expr & \bnfdef &
    x \in X
    \bnfalt u \in  U
    \bnfalt n \in \mathbb{Z}
    \bnfalt b \in \mathbb{B}
    \bnfalt \eiterator{y}
    \\
    &  & &
    \bnfalt \enewlist{\basety}
    \bnfalt \expr_1 \oplus \expr_2
    \bnfalt \expr_1 \bowtie \expr_2
    \bnfalt \expr_1 \lor \expr_2
    \bnfalt \neg \expr
    \\
    Variables & u, x, y, z & \bnfdef &
    x \in X
    \bnfalt u \in U
    \bnfalt y, z \in X \cup U
  \end{tabular}
  }
  \\
  \subfloat[Types and refinements.]
           {
             \label{fig:types-and-refinements}
             \begin{tabular}{ l L L L }
    Base types & \basety & \bnfdef &
    \kwint
    \bnfalt \kwbool
    \bnfalt \titerator{\basety}
    \bnfalt \tlist{\basety}
    \\
    Refinement types & \ty & \bnfdef &
    \trefine{\basety}{\kwrefine}
    \\
    Refinements & \kwrefine & \bnfdef &
    b \in \mathbb{B}
    \bnfalt \sfxbool{x}
    \bnfalt \isiterof{}{\sfxlist{x}}
    \bnfalt \rfexpr_1 \bowtie \rfexpr_2
    \bnfalt \kwrefine_1 \lor \kwrefine_2
    \bnfalt \neg \kwrefine
    \\
    Refinement expressions & \rfexpr & \bnfdef &
    n \in \mathbb{Z}
    \bnfalt \sfxint{\kwself}
    \bnfalt \sfxint{x}
    \bnfalt \len{\sfxlist{\rfexpr}}
    \bnfalt \iter{\sfxiter{\rfexpr}}
    \bnfalt \rfexpr_1 \oplus \rfexpr_2
    \bnfalt \counter \in C
    \\
    List expressions & \sfxlist{\rfexpr} & \bnfdef &
    \sfxlist{\kwself}
    \bnfalt \sfxlist{x}
    \\
    Iterator expressions & \sfxiter{\rfexpr} & \bnfdef &
    \sfxiter{\kwself}
    \bnfalt \sfxiter{x}
    \\
    Typing context & \tyenv & \bnfdef &
    \emp
    \bnfalt \tyenv, x: \ty
  \end{tabular}
           }
           \caption{
             (a) The core calculus for methods manipulating collections.
             The operator $\oparith$
             stands for   arithmetic operators,  while $\opcomp$ stands for
             comparison operators. 
             (b) The types and corresponding refinements. The subscripts in variables
             $\sfxbool{x}, \sfxint{x}, \sfxlist{x}, \sfxiter{x}\in X \cup 
             U$ are used to emphasize their types, $\oplus$ is arithmetic operator restricted to linear
             arithmetic, and $\bowtie$ is a comparison operator.
           }
\vspace{-2em}
\end{figure}

\noindent\textbf{Core calculus.}
Our core calculus focuses on methods manipulating collections as shown
in Figure~\ref{fig:syntax}.
A method $M$ is composed of a sequence of input-variable declarations
$\overline{\edecl{\ty}{u}}$, a sequence of initialized local-variables declarations
$\overline{\edecl{\ty}{\var} = \expr}$, and a method body $s$ that is composed
of basic and compound statements. 
We denote the set of input variables and local variables by $U$ and $X$,
respectively.
The basic statements $x{=}\enext{z}$, $\eremove{y}$, and $\eadd{y}{x}$ provide
standard operations on iterator variable $z$ and collection variable $y$. 
In addition, we have standard assignment statement $x = e$, where $\expr$ is an
expression without side effects. 

\noindent\textbf{Refinement type system.} 
Our type system, shown in Figure~\ref{fig:types-and-refinements},
permits type refinements over base types integer $\kwint$, boolean $\kwbool$,
iterator $\kwiterator$ and list $\kwlist$. 
A refinement type $\trefine{\basety}{\kwrefine}$ further qualifies variables by
providing an assertion over the values of the variable using a predicate
$\kwrefine$.
A unique feature of our refinement predicates is that,
the predicates can refer to AST counters $\counter\in C$ to track limited flow-sensitive information.
Moreover, predicate can refer to the variable on which the refinement is
expressed using the self-reference variable $\kwself$.
A refinement can be expressed as an arbitrary Boolean combination of Boolean
values $b$, Boolean-typed program variables $\sfxbool{x}$, predicates
$\isiterof{}{\sfxlist{x}}$ (expressing that the variable is an iterator of a list
variable $\sfxlist{x}$),  and comparisons between {\it refinement expressions}.
A refinement expression $\rfexpr$ is integer-typed and can be composed of
integer values $n$, integer-typed variable $\sfxint{x}$, length expressions
$\len{\sfxlist{\rfexpr}}$ (representing the length of list expression
$\sfxlist{\rfexpr}$), index expressions $\iter{\sfxiter{\rfexpr}}$ (representing the
current index of an iterator expression $\sfxiter{\rfexpr}$), {\it AST counter
  variables}, and  arithmetic operations over other refinement expressions.
An AST counter variable $\counter\in C$ is associated with an AST node.
Intuitively, it counts the number of times an AST node has been executed.
List expressions $\sfxlist{\rfexpr}$ could be $\sfxlist{\kwself}$ (which refers
to the refined variable itself) or a list-typed program variable $\sfxlist{x}$.
Explanation for the iterator expression $\sfxiter{\rfexpr}$ is analogous.
Typing context $\tyenv$ is a mapping from variables to their types.
Overall, our refinement language is in a decidable logic fragment EUFLIA
(\textoverline{E}quality, \textoverline{U}ninterpreted \textoverline{F}unctions
and \textoverline{LI}near \textoverline{A}rithmetic) where
$\len{\sfxlist{\rfexpr}}$, $\iter{\sfxiter{\rfexpr}}$ and $\isiterof{}{\sfxiter{\rfexpr}}$ are treated as
uninterpreted functions. 

\subsection{Operational semantics}
\label{subsec:dynamics}
\begin{figure}
  \centering
  \subfloat[Environment and Values] {
      \begin{tabular}{ l L L L }
	Environment & \env & \bnfdef &
	\emp
	\bnfalt \mapext{\env}{x}{v}
	\bnfalt \mapext{\env}{u}{v}
	\bnfalt \mapexthook{\env}{\counter}{n}
	\\
	Values & \val & \bnfdef &
	n \in \mathbb{Z}
	\bnfalt b \in \mathbb{B}
	\bnfalt \iterval{n \in \mathbb{N}}{x \in X \cup U}
	\bnfalt \listval{v_1, v_2, \ldots, v_n}
      \end{tabular}
  }
\\
\subfloat[Operational semantics]{
  \begin{mathpar}
    \inferrule* [Lab={\scriptsize E-Var}]
		{
		  {}
		}
		{
		  \evalexpr
		      {\env}
		      {x}
		      {\env[x]}
		}
                \hfill
		\inferrule* [Lab={\scriptsize E-ArithL}]
		            {
			      \evalexpr
			          {\env}
			          {\expr_1}
			          {\expr_1'}
		            }
		{
			\evalexpr
			{\env}
			{\expr_1 {\oparith} \expr_2}
			{\expr_1' {\oparith} \expr_2}
		}                \hfill
		\inferrule* [Lab={\scriptsize E-ArithR}]
		{
			\evalexpr
			{\env}
			{\expr_2}
			{\expr_2'}
		}
		{
			\evalexpr
			{\env}
			{v {\oparith} \expr_2}
			{v {\oparith} \expr_2'}
		}                
                \hfill
                                               \inferrule* [Lab={\scriptsize E-Neg}]
		{
			\evalexpr
			{\env}
			{\expr}
			{\expr'}
		}
		{
			\evalexpr
			{\env}
			{\neg \expr}
			{\neg \expr'}
		}
               	\\
	 \inferrule* [Lab={\scriptsize E-CompL}]
		{
			\evalexpr
			{\env}
			{\expr_1}
			{\expr_1'}
		}
		{
			\evalexpr
			{\env}
			{\expr_1 \bowtie \expr_2}
			{\expr_1' \bowtie \expr_2}
		}\hfill
	        \inferrule* [Lab={\scriptsize E-CompR}]
		{
			\evalexpr
			{\env}
			{\expr_2}
			{\expr_2'}
		}
		{
			\evalexpr
			{\env}
			{v \bowtie \expr_2}
			{v \bowtie \expr_2'}
		}                \hfill
		\inferrule* [Lab={\scriptsize E-OrL}]
		{
			\evalexpr
			{\env}
			{\expr_1}
			{\expr_1'}
		}
		{
			\evalexpr
			{\env}
			{\expr_1 {\lor} \expr_2}
			{\expr_1' \text{ or }  \expr_2}
		}                \hfill
		\inferrule* [Lab={\scriptsize E-OrR}]
		{
			\evalexpr
			{\env}
			{\expr_2}
			{\expr_2'}
		}
		{
			\evalexpr
			{\env}
			{v {\lor} \expr_2}
			{v {\lor} \expr_2'}
		}	
		\\
                {}
                \hfill
                \inferrule * [Lab={\scriptsize E-Counter}]
		{
			\getcnt{\stmt}{M} = \counter
		}
		{
			\evalcounter
			{\env}
			{\stmt}
			{
				\mapexthook
				{\env}
				{\counter}
				{\env[\counter]+1}
			}
		}
                \hfill
                \inferrule * [Lab={\scriptsize E-Counter-Aux}]
		{
			\getcnt{\stmt}{M} = \bot
		}
		{
			\evalcounter
			{\env}
			{\stmt}
			{
				{\env}
			}
		}
                \hfill
\\
{}\hfill
\inferrule* [Lab={\scriptsize E-Iterator}]
		            {
                              {}
		}
		{
			\evalexpr
			{\env}
			{\eiterator{y}}
			{\iterval{0}{y}}
		}                \hfill
		\inferrule* [Lab={\scriptsize E-NewList}]
		            {
                              {}
		}
		{
			\evalexpr
			{\env}
			{\enewlist{\basety}}
			{\listval{}}
		}
                \hfill
      		\inferrule* [Lab={\scriptsize E-Assign}]
		{
			\evalcounter{\env}{x = \expr}{\env'}
			\\
			\evalexprn{\env}{\expr}{v}
		}
		{
			\eval
			{\env}
			{\store}
			{x = \expr}
			{\mapext{\env'}{x}{v}}
			{\store}
			{\eskip}
		}                
\hfill
		\\
		\inferrule * [Lab={\scriptsize E-Next}]
		{
			\evalcounter{\env}{x = \enext{z}}{\env'}
			\\
			\env[z] = \iterval{i}{y}
			\\\\
			\env[y] = \listval{v_1,\ldots,v_n}
			\\
			i < n{-}1
		}
		{
			\eval
			{\env}
			{\store}
			{x = \enext{z}}
			{
				\mapext
				{\mapext{\env'}{z}{\iterval{i+1}{y}}}
				{x}
				{v_{i+1}}
			}
			{\store}
			{\eskip}
		}
		\\
		\inferrule* [Lab={\scriptsize E-Add}]
		{
			\evalcounter{\env}{\eadd{y}{x}}{\env'}
			\\
			\env[y] = \listval{\overline{v}}
			\\
			\evalexpr{\env}{x}{v}
		}
		{
			\eval
			{\env}
			{\store}
			{\eadd{y}{x}}
			{
				\mapext
				{\env'}
				{y}
				{\listappend{\overline{v}}{v}}
			}
			{\store'}
			{\eskip}
		} \                \hfill
		\inferrule* [Lab={\scriptsize E-Remove}]
		{
			\evalcounter{\env}{\eremove{y}}{\env'}
			\\
			\env[y] = \listappend{\overline{v}}{v}
		}
		{
			\eval
			{\env}
			{\store}
			{\eremove{y}}
			{
				\mapext
				{\env'}
				{y}
				{\listval{\overline{v}}}
			}
			{\store'}
			{\eskip}
		}
		\\
		\inferrule * [Lab={\scriptsize E-IfExpr}]
		{
			\evalcounter{\env}{\eif{\expr}{\stmt_1}{\stmt_2}}{\env'}
			\\
			\evalexpr{\env}{\expr}{\expr'}
		}
		{
			\eval
			{\env}
			{\store}
			{\eif{\expr}{\stmt_1}{\stmt_2}}
			{\env'}
			{\store}
			{\newcnt\eif{\expr'}{\stmt_1}{\stmt_2}}
		}
                \\
		\inferrule * [Lab={\scriptsize E-IfTrue}]
		{
			\evalcounter{\env}{\eif{\kwtrue}{\stmt_1}{\stmt_2}}{\env'}
		}
		{
			\eval
			{\env}
			{\store}
			{\eif{\kwtrue}{\stmt_1}{\stmt_2}}
			{\env'}
			{\store}
			{\stmt_1}
		}
		\hfill
		\inferrule * [Lab={\scriptsize E-IfFalse}]
		{
			\evalcounter{\env}{\eif{\kwfalse}{\stmt_1}{\stmt_2}}{\env'}
		}
		{
			\eval
			{\env}
			{\store}
			{\eif{\kwfalse}{\stmt_1}{\stmt_2}}
			{\env'}
			{\store}
			{\stmt_2}
		}                \\
		\inferrule * [Lab={\scriptsize E-While}]
		{
			\evalcounter{\env}{\ewhile{\expr}{\stmt}}{\env'}
		}
		{
			\eval
			{\env}
			{\store}
			{\ewhile{\expr}{\stmt}}
			{\env'}
			{\store}
			{
				\newcnt\eif{\expr}
				{
					\newcnt\{\stmt;\newcnt\ewhile{\expr}{\stmt}\}
				}
				{\eskip}
			}
		}
		\\
		\inferrule * [Lab={\scriptsize E-Block}]
		{
			\evalcounter{\env}{\eblock{\stmt}}{\env'}
			\\
			\overline{\stmt} = \stmt_1;\stmt_2;\ldots;\stmt_n
			\\\\
			\eval{\env'}{\store}{\stmt_1}{\env''}{\store'}{\stmt_1'}
			\\
			\overline{\stmt'} = \newcnt{\stmt_1'};\stmt_2;\ldots;\stmt_n
		}
		{
			\eval
			{\env}
			{\store}
			{\eblock{\stmt}}
			{\env''}
			{\store'}
			{\newcnt\eblock{\stmt'}}
		}                \hfill
		\inferrule * [Lab={\scriptsize E-BlockSkip}]
		{
			\overline{\stmt} = \eskip;\stmt_2;\ldots;\stmt_n
			\\
			\overline{\stmt'} = \stmt_2;\ldots;\stmt_n
		}
		{
			\eval
			{\env}
			{\store}
			{\eblock{\stmt}}
			{\env}
			{\store}
			{\newcnt{\eblock{\stmt'}}}
		}
    \end{mathpar}
}
  \caption{Environment,values, and small-step operational semantics.}
  \label{fig:operational-semantic}
\end{figure}

We define small-step operational semantics of our core calculus as well as
semantics of type refinements in Figure \ref{fig:operational-semantic} and \ref{fig:type-semantic}.
An environment (or equivalently, a state) $\env$ is a mapping from program variables to values, which
intuitively serves as a stack activation record.
The domain of variable values include integers, booleans, iterators, and list values.
The calculus also supports lists of lists.
We denote the initial environment as $\env_\text{init}$.
Environment $\env_\text{init}$ initializes counters as zero, input variables as
concrete input values, and local variables as their initial values specified in
the method. 

Figure \ref{fig:operational-semantic} defines the small-step operational semantics
for our core calculus.
We use the following three judgment forms:
\begin{enumerate}
\item Judgment form $\evalexpr{\env}{\expr}{\expr'}$ states that expression $\expr$
  is evaluated to expression $\expr'$ in one evaluation step under environment
  $\env$,
\item Judgment form $\eval{\env}{}{\stmt}{\env'}{}{\stmt'}$ states that
  after one evaluation step of executing statement $\stmt$ under environment
  $\env$, the environment changes to $\env'$ and the next statement to be
  evaluated is $\stmt'$, and
\item Judgment form $\evalcounter{\env}{\stmt}{\env'}$ expresses the AST counter state
  transitions by modifying $\env$ to increment the counter
  value associated with statement $\stmt$.
\end{enumerate}
Compared with standard operational semantics (IMP language
\cite{Win93}), there are two main differences. 
The first difference is that we introduce collections into our core calculus. 
The semantics of collection operations is straightforward as shown in
Figure~\ref{fig:operational-semantic}. 
The other significant difference is due to the use of AST counters in refinement types. Most of
the differences from non-standard semantics is related to handling of these
counters.
The function $\getcnt{\stmt}{M}$ returns the unique counter $\counter$
associated with the statement $\stmt$ in the method $M$.
Notice that the intermediate derivations of the rules may produce auxiliary statements
that are not present in the original program.
Since the refinement types may not refer to these counters, we ignore counter
values for these auxiliary statements by associating them with a same special counter $\bot$, whose value we do not care about.
E.g., The conclusion of the rule $\textsc{E-IfExpr}$
introduces a new $\newcnt\texttt{if}$ statement along with original statements $s_1$ and $s_2$, associating this new if-else statement with counter $\bot$.
Rules \textsc{E-Counter} and \textsc{E-Counter-Aux} are mainly concerned with
AST counter bookkeeping.
The explanation of other rules is straightforward.

\noindent\textbf{Types and Refinements.} Figure \ref{fig:type-semantic} defines
semantics of types and refinements.
Judgment form $\evalty{\val}{\env}{}{\ty}$ states that the value $\val$ conforms
to a type $\ty$ under environment $\env$.
The semantics of the base-types $\evalty{\env[x]}{\env}{}{\basety}$ is
straightforward and hence omitted. 
The judgment form $\evalrefine{\rftuple{\kwrefine}{x}}{\env}$ states that
variable $x$ to which expression $\kwself$ in refinement $\kwrefine$ refers, 
conforms to the refinement under the environment $\env$. 
We exploit helper
functions $\evalrfexpr{\rftuple{\rfexpr}{x}}{\env}$ and 
$\unintsubst{\rftuple{\rfexpr}{x}}{\env}$ in refinement semantics defined in the following fashion:
\begin{itemize}
\item Function $\evalrfexpr{\rftuple{\rfexpr}{x}}{\env}$ takes a refinement
  expression $\rfexpr$, a variable $x$ (to which self-reference $\kwself$ in
  $\rfexpr$ refers), and an environment $\env$ as inputs and
  then returns the evaluation of refinement expression.
  
\item Function $\unintsubst{\rftuple{\rfexpr}{x}}{\env}$ takes as inputs
  refinement expression $\rfexpr$, variable $x$ (to which expression $\kwself$
  refers), and environment $\env$, and returns an expression that is the
  result of first substituting self-reference $\kwself$ with variable $x$ and then
  substituting every $\len{\sfxlist{x}}$ in $\rfexpr$ with length of list-typed
  variable $\sfxlist{x}$, as well as every $\iter{\sfxiter{y}}$ with index value
  of iterator-typed variable $\sfxiter{y}$. 
\end{itemize}
We write $\symtrans^{*}$ for the transitive closure of $\symtrans$.
Most of the refinement semantics are straightforward.
In particular, the semantics of $\isiterof{}{y}$ is that variable $x$, to which
$\kwself$ refers, is an iterator for list-typed variable $y$.

\begin{figure}[t]
  \centering
  \begin{tabular}{ L p{0.5cm} L }
    \evalty{\env[x]}{\env}{}{\trefine{\basety}{\kwrefine}} & iff &
    \evalty{\env[x]}{\env}{}{\basety} \text{ and }
    \evalrefine{\rftuple{\kwrefine}{x}}{\env}
    \\
    \evalrefine{b}{\env} & iff &
    b \equiv \kwtrue
    \\
    \evalrefine{\rftuple{\sfxbool{y}}{x}}{\env} & iff &     \evalrefine{\env[\sfxbool{y}]}{\env} 
    \\
    \evalrefine{\rftuple{\neg \kwrefine}{x}}{\env} & iff &
    (\evalrefine{\rftuple{\kwrefine}{x}}{\env}) \not = \kwtrue
    \\
    \evalrefine{\rftuple{\kwrefine_1\lor\kwrefine_2}{x}}{\env} & iff &
    \evalrefine{\rftuple{\kwrefine_1}{x}}{\env} \text{ or } \evalrefine{\rftuple{\kwrefine_2}{x}}{\env}
    \\
    \evalrefine{\rftuple{\isiterof{}{y}}{x}}{\env} & iff &
    \text{ for some $i \geq 0$ we have }\env[x] = \iterval{i}{y}
    \\
    \evalrefine{\rftuple{\rfexpr_1\opcomp\rfexpr_2}{x}}{\env} & iff &
    \evalrfexpr{\rftuple{\rfexpr_1}{x}}{\env} \bowtie \evalrfexpr{\rftuple{\rfexpr_2}{x}}{\env}
    \\
    \evalrfexpr{\rftuple{\rfexpr}{x}}{\env} & = &
    v
    \text{, where }
    e = \unintsubst{\rftuple{\rfexpr}{x}}{\env} \land
    \evalexprn{\env}{e}{v}
    \\
    \unintsubst{\rftuple{\rfexpr}{x}}{\env} & = &
    (\rfexpr[x/\kwself])[n_i/\len{\sfxlist{x^i}},k_j/\iter{\sfxiter{y^j}}]_{
    \text{ $\forall \sfxlist{x^i}, \sfxiter{y^j}$}}
    \\ & &
    \text{ where } \env[\sfxlist{x^i}] = \listval{v_1, \ldots, v_{n_i}}
    \text{ and } \env[\sfxiter{y^j}] = \iterval{k_j}{*}
  \end{tabular}
  \caption{Refinement semantics.}
  \label{fig:type-semantic}
  \vspace{-2em}
\end{figure}

\subsection{Well-Typed Methods}
\label{subsec:well-typed-methods}
We say that an environment $\env$ is {\it reachable} in a method $M$ if
$\evaln{\env_\text{init}}{}{M}{\env}{}{\stmt}$.
We write $\Reach(M)$ for the set of all reachable environments of $M$.
We say that an environment $\env$ is {\it well-typed} in $M$ if all of the
variables conform to their types, i.e. for all $x \in X \cup U$ with type $\trefine{\basety}{\kwrefine}$, we have that
$\evalty{\env[x]}{\env}{}{\trefine{\basety}{\kwrefine}}$.
We write $\WellTyped(M)$ for the set of all well-typed environments in $M$.
We say that a method $M$ is {\it well-typed} if all of the reachable states
of $M$ are well-typed, i.e. $\Reach(M) \subseteq \WellTyped(M)$. 

  

\vspace{-1em}
\section{Collection Bound Verification Problem}
\label{sec:algorithm}
Given a method $M$, our goal is to verify that if the inputs to the method
satisfy a given assumption $\phi_\Aa$, then the method $M$ guarantees that the
collection lengths remain bounded. 
The guarantee requirements $\phi_\Gg$ can be expressed as a predicate
constructed using the refinement language introduced in Figure
\ref{fig:types-and-refinements}.
Observe that, since this verification condition is not attached to any
particular variable, it is free from predicates $\isiterof{}{\sfxlist{x}}$ and
self-reference $\kwself$.
We further assume that the assumptions on the input variables are expressed
using type refinements on the input variables.
Formally, we are interested in the following problem:
\begin{definition}[Collection Bound Verification Problem]
  \label{definition:problem-statement}
  Given a method $M$ along with its input variables with types and refinements 
  $u_i:\ty_i$, and a guarantee requirement $\phi_\Gg$, verify that every
  reachable environment satisfies $\phi_\Gg$, i.e. for all
  $\env \in \Reach(M)$
  we have that $\evalrefine{\phi_\Gg}{\env}$.
\end{definition}

We present  a type-directed approach to solve this problem.
We first propose type-checking rules to verify if the method is well-typed. 
Then, we discuss how to automatically derive AST counter relation axioms in
Section~\ref{subsec:relational-control-flow-abstraction}. 
Finally, we reduce solving the verification problem into issuing SMT queries, in Section
\ref{subsec:bound-checking}, using as
constraints the type refinements verified in
Section \ref{subsec:type-checking} as well as AST counter relation axioms
extracted from Section \ref{subsec:relational-control-flow-abstraction} .

\subsection{Type checking}
\label{subsec:type-checking}
\begin{figure}
  \centering
  \begin{mathpar}
    \inferrule* [Lab={\scriptsize T-Add}]
		{
		  \getcnt{\eadd{y}{x}}{M} = \counter
		  \\
		  \forall (w: \ty_w) \in \tyenv.
		  \subtyof
		      {(\substitute{\ty_w}{[w/\kwself]})}
		      {
				\substitute
				{(\substitute{\ty_w}{[w/\kwself]})}
				{
				  [(\len{y}{+}1)/\len{y}, (\counter{+}1)/\counter]
				}
			}
		}
		{
			\tycheck
			{\tyenv}
			{\eadd{y}{x}}
		}
		\\
		\inferrule* [Lab={\scriptsize T-Remove}]
		{
			\getcnt{\eremove{y}}{M} = \counter
			\\
			\forall (w: \ty_w) \in \tyenv.
			\subtyof
			{(\substitute{\ty_w}{[w/\kwself]})}
			{
				\substitute
				{(\substitute{\ty_w}{[w/\kwself]})}
				{
                                  [(\len{y}{-}1)/\len{y}, (\counter{+}1)/\counter]
				}
			}
		}
		{
			\tycheck
			{\tyenv}
			{\eremove{y}}
		}
		\\
		\inferrule* [Lab={\scriptsize T-AssignIter}]
		{
			\getcnt{z = \eiterator{y}}{M} = \counter
			\\
			\forall (w: \ty_w) \in \tyenv.
			\subtyof
			{(\substitute{\ty_w}{[w/\kwself]})}
			{
				\substitute
				{(\substitute{\ty_w}{[w/\kwself]})}
				{[0/\iter{z},(\counter{+}1)/\counter]}
			}
			\\
			\tycheck{\tyenv}{z: \ty_z}
			\\
			\subtyof
			{(\substitute{\ty_z}{[z/\kwself]})}
			{
				\substitute
				{(\substitute{\ty_z}{[z/\kwself]})}
				{
					[0/\iter{z} ,(\counter{+}1)/ \counter,\isiterof{z}{y}/\isiterof{}{*}]
				}
			}
		}
		{
			\tycheck
			{\tyenv}
			{z = \eiterator{y}}
		}
		\\
		\inferrule* [Lab={\scriptsize T-Assign}]
		{
			\getcnt{x = e}{M} = \counter
			\\
			\tycheck{\tyenv}{x: \trefine{\basety}{\kwrefine_x}}
			\\
			\basety \text{ is not a list type}
			\\\\
			\forall (w: \ty_w) \in \tyenv.
			\subtyof
			{(\substitute{\ty_w}{[w/\kwself]})}
			{
				\substitute
				{(\substitute{\ty_w}{[w/\kwself]})}
				{
					[e/x,(\counter{+}1)/\counter]
				}
			}
		}
		{
			\tycheck
			{\tyenv}
			{x = e}
		}
		\\
		\inferrule* [Lab={\scriptsize T-AssignList}]
		{
			\getcnt{x = e}{M} = \counter
			\\
			\tycheck{\tyenv}{x: \trefine{\basety}{\kwrefine_x}}
			\\
			\basety \text{ is a list type}
			\\\\
			\forall (w: \ty_w) \in \tyenv.
			\subtyof
			{(\substitute{\ty_w}{[w/\kwself]})}
			{
				\substitute
				{(\substitute{\ty_w}{[w/\kwself]})}
				{
					[\len{e}/\len{x},(\counter{+}1)/\counter]
				}
			}
		}
		{
			\tycheck
			{\tyenv}
			{x = e}
		}
		\\
		\inferrule* [Lab={\scriptsize T-AssignNewList}]
		{
			\getcnt{x = \enewlist{\basety}}{M} = \counter
			\\\\
			\forall (w: \ty_w) \in \tyenv.
			\subtyof
			{(\substitute{\ty_w}{[w/\kwself]})}
			{
				\substitute
				{(\substitute{\ty_w}{[w/\kwself]})}
				{
					[0/\len{x}, (\counter{+}1)/\counter]
				}
			}
		}
		{
			\tycheck
			{\tyenv}
			{x = \enewlist{\basety}}
		} \\
		\inferrule* [Lab={\scriptsize T-Counter}]
		{
			\stmt \ne \stmt_B
			\\
			\getcnt{\stmt}{M} = \counter
			\\\\
			\forall (w: \ty_w) \in \tyenv.
			\subtyof
			{(\substitute{\ty_w}{[w/\kwself]})}
			{
				(\substitute{\ty_w}{[w/\kwself]})
				[(\counter{+}1)/\counter]
			}
		}
		{
			\tycheckcnt
			{\tyenv}
			{\stmt}
		}
		\\
		\inferrule* [Lab={\scriptsize T-Next}]
		{
			\getcnt{x = \enext{z}}{M} = \counter
			\\\\
			\forall (w: \ty_w) \in \tyenv.
			\subtyof
			{(\substitute{\ty_w}{[w/\kwself]})}
			{
				\substitute
				{(\substitute{\ty_w}{[w/\kwself]})}
				{
					[(\iter{z}{+}1)/\iter{z}, (\counter{+}1)/\counter]
				}
			}
			\\\\
			\forall (w: \trefine{\ty_w}{\kwrefine_w}) \in \tyenv.
			\kwrefine_w \Longleftrightarrow \exists x.\kwrefine_w 
		}
		{
			\tycheck
			{\tyenv}
			{x = \enext{z}}
		}\\
		\inferrule* [Lab={\scriptsize T-Decl}]
		{
		  \evalexpr{\env_\text{init}}{e}{v}
                  \\
                  \tycheck{\tyenv}{\stmt}
		  \\
		  \forall \subastnode{\edecl{\ty}{x} = e}{M}. \evalty{v}{\env_\text{init}}{}{\ty}
		}
		{
			\tycheck
			{\tyenv}
			{
				\overline{\edecl{\ty}{u}}\;
				\overline{\edecl{\ty}{x} = e}\;
				\stmt
			}
		}
		\inferrule* [Lab={\scriptsize T-Block}]
		{
			\overline{\stmt} = \stmt_1;\ldots;\stmt_n
			\\\\
			\tycheckcnt{\tyenv}{\eblock{\stmt}}
			\\\\
			{\tycheck{\tyenv}{\stmt_i}}
			\text{, for all } i \in \{1,\cdots,n \}
		}
		{
			\tycheck{\tyenv}{\eblock{s}}
		}\\
		\inferrule* [Lab={\scriptsize T-While}]
		{
			\tycheckcnt{\tyenv}{\ewhile{\expr}{\stmt}}
			\\
			\tycheck{\tyenv}{\stmt}
		}
		{
			\tycheck{\tyenv}{\ewhile{\expr}{\stmt}}
		}
		\inferrule* [Lab={\scriptsize T-Skip}]
		{
			{}
		}
		{
			\tycheck{\tyenv}{\eskip}
		}
		\inferrule* [Lab={\scriptsize T-If}]
		{
			\tycheckcnt{\tyenv}{\eif{\expr}{\stmt_1}{\stmt_2}}
			\\
			\tycheck{\tyenv}{\stmt_1}
			\\
			\tycheck{\tyenv}{\stmt_2}
		}
		{
			\tycheck{\tyenv}{\eif{\expr}{\stmt_1}{\stmt_2}}
		}
  \end{mathpar}
  \caption{Type checking rules}
  \label{fig:type-checking-rules}
\end{figure}

Our key analysis algorithm is encoded into refinement type checking rules shown
in Figure~\ref{fig:type-checking-rules}. 
Subtyping between two refinement types is defined as the implication relation
between two refinements using the following rule:
\begin{mathpar}
  \inferrule*[Right={\scriptsize <:-RefinementTyp}]  {
    \subtyof{\basety_1}{\basety_2} \hfill
    \\
    \kwrefine_1 \implies \kwrefine_2 }{
    \subtyof{\trefine{\basety_1}{\kwrefine_1}}{\trefine{\basety_2}{\kwrefine_2}}
  }
\end{mathpar}

Figure \ref{fig:type-checking-rules} defines type-checking rules for refinement
types, while the rules for base types are standard and thus presented in
Appendix~\ref{sec:base-type-checking}.
Notation $\substitute{\ty}{[{\rfexpr}'/\rfexpr]}$ denotes substituting expression $\rfexpr$ with ${\rfexpr}'$ in the refinement of type $\ty$.

The Judgment form $\tycheck{\tyenv}{\stmt}$ states that the statement $\stmt$ is
successfully type checked under typing context $\tyenv$ if premises are
satisfied. 
We case split on the right hand side of assignment statement $x=e$ into: Rule
\textsc{T-AssignIter}, \textsc{T-Assign}, \textsc{T-AssignList}, and
\textsc{T-AssignNewList}. 
Intuitively, type checking rules check that after applying each corresponding
evaluation rule, type refinements should still be valid. 
More specifically, in each type checking rule we check for all refinements, if
its validity before applying a corresponding evaluation rule implies its
validity afterwards. 
For example, after applying Rule \textsc{E-Add}, the environment has the
following updates: 
length of collection variable $y$ is incremented by 1 and
the associated AST counter's value is incremented by 1.
Therefore Rule \textsc{T-Add} checks the implication of validity between a type
$\substitute{\ty_w}{[w/\kwself]}$ and the result after applying to it a
substitution
$\substitute{(\substitute{\ty_w}{[w/\kwself]})}{[(\len{y}{+}1)/\len{y}, (\counter{+}1)/\counter]}$,
which precisely expresses the actual value of type
$\substitute{\ty_w}{[w/\kwself]}$ after applying Rule \textsc{E-Add} in terms of
its value beforehand. 
Rule \textsc{T-Remove} is dual to Rule \textsc{T-Add}.
In Rule \textsc{T-AssignIter}, in addition to subtyping checking, we also check
for variable $z$ if its refinement will still hold true after substituting
$\isiterof{}{*}$ with $\isiterof{}{y}$. 
The intuition behind is that after evaluating statement $z=\eiterator{y}$, variable $z$
will become an iterator for variable $y$, no matter what list it was an iterato for.
For a reader interested in why we must treat refinement $\isiterof{}{x}$ differently, the root cause here is that unlike $\iter{z}$ specifying a property of \textbf{one} variable, $\isiterof{}{x}$ actually specifies a relation between \textbf{two} variables.
Rule \textsc{T-Assign} checks if refinements will still hold true when $x$
becomes $e$, no matter if variable $x$ is integer-typed, boolean-typed or
iterator-typed (where $\iter{x}$ becomes $\iter{e}$). 
Rule \textsc{T-AssignList} and Rule \textsc{T-Assign} are similar,
except that in Rule \textsc{T-AssignList} we check if refinements will still
hold true when $\len{x}$ becomes $\len{e}$. 
We split Rule \textsc{T-AssignList} from Rule Rule \textsc{T-Assign}, avoiding
simply checking if $x$ becoming $e$ will break any refinement, because assignment
$x=e$ does not make refinement $\isiterof{}{x}$ become $\isiterof{}{e}$.
In Rule \textsc{T-Next}, besides checking the validity of implication, we also check if every type refinement is logically equivalent to itself being existentially quantified by variable $x$.
Intuitively, this ensures soundly that the assignment in statement $x=\enext{z}$ will not break any refinement, since there is no constraint on list elements retrieved from list variable $z$ by invoking $\enext{z}$.
Just like Rule \textsc{E-Counter} interleaves with every evaluation rule in
Figure \ref{fig:operational-semantic}, Rule \textsc{T-Counter} serves as
a premise for every type checking rule of compound statements. 
For every type checking rule of basic statements, Rule \textsc{T-Counter} is
embedded into subtyping checking. 
Rule \textsc{T-Decl} checks that all local variables' type refinements are
valid, given their initial values. 
We also define a helper function $\subastnode{\stmt_1}{\stmt_2}$ that is used in Rule \textsc{T-Decl}, which describes AST
sub-node relations between AST node $\stmt_2$ and its sub-node $\stmt_1$.
\begin{mathpar}
  \hfill
  \inferrule* [Lab={SubNode-Block}] {
    \overline{\stmt} = \stmt_1;\ldots;\stmt_n
  } {
    \subastnode
	{{\stmt_i}}
	{\eblock{\stmt}},
   \text{ for all $i \in \{1,\cdots,n\}$}
}\hfill
  \inferrule* [Lab={SubNode-While}]{
    {}
  }{
    \subastnode{\stmt}{\ewhile{e}{\stmt}}
  }\hfill\\
  \inferrule* [Lab={SubNode-If}]
	      {
	      }
	      {
		\subastnode
		    {{\stmt_i}}
		    {\eif{e}{\stmt_1}{\stmt_2}}
                    , \text{ for $i \in \{1,2\}$}
	      }
              
\end{mathpar}

\subsection{AST Counter Axioms}
\label{subsec:relational-control-flow-abstraction}
\begin{figure}[t]
  \begin{mathpar}
    \inferrule* [Lab={\scriptsize R-Block}]
		{
		  \overline{\stmt} = \stmt_1;\ldots;\stmt_n
		  \\
		 \rcfa{\stmt_i} = C_i \text{ and } \getcnt{\stmt_i}{M} = \counter_i
		 \text{, for all } i \in \{1,\cdots,n\}
		 \\\\
		  d_j = 
		  (
		  \bigwedge\limits_{i=1}^{j}
		  \counter_i \equiv \counter_{i-1} \land
		  \counter_{j{+}1}{+}1 \equiv \counter_j
		  )
		  \text{, for all } j \in \{1,\cdots,n{-}1\}
		  \\\\
		  \getcnt{\{\overline{\stmt}\}}{M} = \counter_0
		  \\
		  d_n = (\counter_0\equiv\cdots\equiv\counter_n)
		}
		{
		  \rcfa{\eblock{\stmt}} =
		  \bigwedge\limits_{i=1}^n C_i
		  \land
		  \bigvee\limits_{i=1}^n d_i
		}\\
                {}
                \hfill
		\inferrule* [Lab={\scriptsize R-While}]
		            {
			      \rcfa{\stmt} = C
			\\
			\getcnt{\ewhile{\expr}{\stmt}}{M} = \counter_0
			\\
			\getcnt{\stmt}{M} = \counter_\text{b}
		}
		{
			\rcfa{\ewhile{\expr}{\stmt}} =
			C
		}
		\hfill
		\inferrule* [Lab={\scriptsize R-Basic}]
		{
			{}
		}
		{
			\rcfa{\stmt_\text{B}} = \kwtrue
		}\hfill\\
		\inferrule* [Lab={\scriptsize R-If}]
		{
			\rcfa{\stmt_i} = C_i \text{ and } \getcnt{\stmt_i}{M} = \counter_i
			\text{, for } i \in \{1,2\}
			\\
			\getcnt{\eif{\expr}{\stmt_1}{\stmt_2}}{M} = \counter_0
		}
		{
			\rcfa{\eif{\expr}{\stmt_1}{\stmt_2}} =
			(\counter_0 \equiv \counter_1 + \counter_2) \land C_1 \land C_2
		}
	\end{mathpar}
	\caption{AST Counter Axioms}
	\label{fig:relational-control-flow-abstraction}
	\vspace{-2em}
\end{figure}

We next present the AST counter relation axioms.
The goal of deriving counter relation axioms is to improve
verification precision by having additional constraints when encoding the
problem statement into SMT queries. 
We let counter relations precisely correspond to abstract syntax tree
structure of a program.
Respecting semantics of counters, these counters keep record of the number of
times a particular AST node has been executed at runtime.

The function $\rcfa{\stmt}$ takes as input a statement $\stmt$ and
statically outputs a predicate about the relations on all AST 
sub nodes of statement $\stmt$, as well as counter relation axioms derived from all AST sub nodes themselves.
For example, Rule \textsc{R-Block} extracts counter relations from a block of
statements $\eblock{\stmt}$. 
For $1 \leq j \leq n-1$, in the constraint $d_j$ the counter $\counter_i$
associated with statement $\stmt_i$ is either: a) equal to counter
$\counter_{i+1}$ associated with statement $\stmt_{i+1}$, when 
statement $\stmt_i$ and $\stmt_{i+1}$ have both been executed; or b) the counter
$\counter_i$ is equal to $\counter_{i+1}{+}1$, when statement $\stmt_i$ 
has been executed, but not statement $\stmt_{i+1}$. 
Intuitively, constraint $d_j$ describes a set of valid counter relations at one
program state, which is immediately after executing statement $\stmt_j$ but
before executing statement $\stmt_{j+1}$. 
Constraint $d_n$ denotes the counter relations right after finishing executing block statement $\eblock{\stmt}$.
Additionally, the value of counter $\counter_0$ (associated with block
statement $\eblock{\stmt}$ itself) is always equivalent to the value of counter
$\counter_1$ (associated with the first statement $\stmt_1$ in the block),
respecting operational semantics of $\eblock{\stmt}$ defined in Rule
\textsc{E-Block} of Figure \ref{fig:operational-semantic}. 
Furthermore, the constraints ${C_i}$, for $1 \leq i \leq n$, are recursively
generated from every statement $\stmt_i$. 
Intuitively, these relations describes counter relations when
flow-sensitively executing the code block $\eblock{\stmt}$.

As another example, Rule \textsc{R-While} extracts counter relations from a
while loop. 
Note that we cannot conclude any relations between counter $\counter_\text{b}$ (associated with loop body
$\stmt$) and counter $\counter_0$ (associated with loop
$\ewhile{\expr}{\stmt}$), because although loop body $\stmt$ may be executed for a positive number times or may not be executed, loop $\ewhile{\expr}{\stmt}$ will always be executed for one more time whenever executing this AST node, according to Rule \textsc{E-While} in Figure \ref{fig:operational-semantic}.
Other rules are straightforward.
Proof of soundness for above counter relations is straightforward and hence omitted.

\vspace{-1em}

\subsection{Collection Bound Verification}
\label{subsec:bound-checking}
We formalize our approach that solves the collection bound
verification problem for method $M$ by constructing an SMT query. 
We first obtain constraints from type refinements and AST counter axioms, and
then generate the following SMT query that searches for counterexamples for the
guarantee $\phi_\Gg$:
\begin{align*}
  \cstrast \land
  \bigwedge\limits_{\subastnode{\edecl{\ty}{u}}{M}} \genconsty{u: \ty} \land
  \bigwedge\limits_{\subastnode{\edecl{\ty}{x}}{M}} \genconsty{x: \ty}
  \land 
  \neg \phi_\Gg,
\end{align*}
where   $\cstrast$ are the constraints generated from functions $\rcfa{s}$
defined in Section~\ref{subsec:relational-control-flow-abstraction}.
The helper function $\genconsty{x:\ty}$, defined in Figure~\ref{helper}, takes
as input a variable $x$ together with its type $\ty$, and returns refinement constraints from type $\ty$. 
\begin{figure}[t]
\begin{center}
    \begin{tabular}{ L L L }
      \genconsty{x:\basety} & \kwdefine & \kwtrue
      \\
      \genconsty{x:\trefine{\basety}{x_\text{bool}}} & \kwdefine & x_\text{bool}[x/\kwself]
      \\
      \genconsty{x:\trefine{\basety}{\rfexpr_1\opcomp\rfexpr_2}} & \kwdefine & \rfexpr_1[x/\kwself] \opcomp \rfexpr_2[x/\kwself]
      \\
      \genconsty{x:\trefine{\basety}{\kwrefine_1\lor\kwrefine_2}} & \kwdefine & \kwrefine_1[x/\kwself]\lor\kwrefine_2[x/\kwself]
      \\
      \genconsty{x:\trefine{\basety}{\neg\kwrefine}} & \kwdefine & \neg\kwrefine[x/\kwself]
      \\
      \genconsty{x:\trefine{\basety}{\isiterof{}{y}}} & \kwdefine & 0 \le \iter{x} \le \len{y}
    \end{tabular}
\end{center}
\caption{The helper function $\Phi$ for extracting refinement constraints.}
\label{helper}
\vspace{-2em}
\end{figure}
Intuitively, constraint $\cstrast$ soundly constrains the possible values that
AST counters could take when flow-sensitively executing a program. 
Constraints $\genconsty{u: \ty}$ encode assumptions on the inputs to the method, and
constraints $\genconsty{x: \ty}$ soundly constrain the values that local variables could
take. 
Together they constitute a constraint on all reachable program states (which is
proven in Section \ref{sec:soundness}).
In other words, the conjunction of constraints defines a set of program states
that is a sound over-approximation of every actual reachable program states of
method $M$. 
Therefore, the answer to the query provides a sound solution to the collection
bound verification problem.

\vspace{-1em}

\section{Soundness}
\label{sec:soundness}
In this section, we present theorems on refinement preservation and refinement
progress.
Intuitively, refinement preservation guarantees that if a program passes
refinement type checking (Section \ref{subsec:type-checking}), then it will
always end up in a well-typed environment (Section
\ref{subsec:well-typed-methods}), under which we perform bound verification
(Section \ref{subsec:bound-checking}). 
Refinement progress states that a program that passes type checking will not get
stuck. 
Refinement preservation is the core theorem, we prove it below.

\begin{theorem}[Refinement preservation]
	\label{theorem:refinement-preservation}
	If we have that $\wellsize{\tyenv}{\env}$, $\tycheck{\tyenv}{\stmt}$, and $\eval{\env}{\store}{\stmt}{\env'}{\store'}{\stmt'}$,
	then $\wellsize{\tyenv}{\env'}$ and $\tycheck{\tyenv}{\stmt'}$.
\end{theorem}
\begin{proof}
  Given $\wellsize{\tyenv}{\env}$ and
  $\tycheck{\tyenv}{\stmt}$ and
  $\eval{\env}{\store}{\stmt}{\env'}{\store'}{\stmt'}$, we focus on proving
  $\wellsize{\tyenv}{\env'}$, because the validity of $\tycheck{\tyenv}{\stmt'}$
  is directly implied from the premises in Figure
  \ref{fig:type-checking-rules}. 
    The goal is to prove for every variable $x_i$ with type $\ty_i$ in
  $\domof{\env}$, we have $\evalty{\env'[x_i]}{\env'}{}{\ty_i[x_i/\kwself]}$. 
  
  \begin{itemize}
  \item \textbf{Rule \textsc{E-Add}}:
    We need to prove that if $\wellsize{\tyenv}{\env}$ and $\eval{\env}{\store}{\eadd{y}{x}}{\env'}{\store'}{\eskip}$, then
    $\wellsize{\tyenv}{\env'}$.
    From the Rule~\textsc{T-Add}, we have \\
    (Fact 1): $\wellsize{\tyenv}{\env}$ implies
    that  $\wellsize{\tyenv[(\len{y}{+}1)/\len{y}, (\counter{+}1)/\counter]}{\env}$, where we define $\tyenv[(\len{y}{+}1)/\len{y}, (\counter{+}1)/\counter]$ as performing substitution $[(\len{y}{+}1)/\len{y}, (\counter{+}1)/\counter]$ for all types in typing context $\tyenv$.
    \\
    From the Rule~\textsc{E-Add}, we can infer that if
    $\eval{\env}{\store}{\eadd{y}{x}}{\env'}{\store'}{\eskip}$, then
    $\env'(z)=\env(z)$ for variables other than $\counter$ and $y$. Furthermore,
    $\env'[\counter]=\env[\counter]+1$ an
    $\len{\env'[y]}=\len{\env[y]}+1$.
    Based on these properties of $\env'$, we prove by a simple induction on the structure of refinements that \\
    (Fact 2): if $\wellsize{\tyenv[(\len{y}{+}1)/\len{y}, (\counter{+}1)/\counter]}{\env}$ then $\tycheck{\env'}{\tyenv}$. \\
    By chaining Fact 1 and Fact 2, we can conclude the proof.
  \end{itemize}
  The other cases are similar or simpler, and can be found in the Appendix \ref{sec:soundness-proof}.
  \qed
\end{proof}

Corollary \ref{corollary:well-typedness} states that all reachable program states are well-typed (Section \ref{subsec:well-typed-methods}).
The proof immediately follows from Theorem \ref{theorem:refinement-preservation}.

\begin{corollary}
	\label{corollary:well-typedness}
	If $\tycheck{\tyenv}{M}$ and
	$\env \in \Reach(M)$
	then $\wellsize{\tyenv}{\env}$.
\end{corollary}


\begin{theorem}[Refinement progress]
	\label{theorem:refinement-safety}
	If $\wellsize{\tyenv}{\env}{}$ and $\tycheck{\tyenv}{\stmt}$,
	then either statement $\stmt$ is $\eskip$,
	or there exist $\env'$ and $\stmt'$ such that $\eval{\env}{\store}{\stmt}{\env'}{\store'}{\stmt'}$
\end{theorem}

The proof of Theorem \ref{theorem:refinement-safety} is standard and hence omitted.

\vspace{-1em}
\section{Experiment}
\label{sec:experiment}

We implemented our tool \toolname\;in Scala using the Checker Framework \cite{dietl2011building,papi2008practical}, Microsoft
Z3 \cite{de2008z3} and Scala SMT-LIB \cite{scalasmtlib}.
\toolname\;is implemented as a Java annotation processor, relying on the Checker Framework to extract type annotations and perform type checking.
Microsoft Z3 served as an off-the-shelf SMT solver.
We also used Scala SMT-LIB for parsing string-typed annotations.
We chose several web applications as benchmarks (180k lines of code in total),
each of which supports various functionalities.
Benchmarks were collected from
different sources, including GitHub (jforum3 with 218 stars and SpringPetClinic
with 2325 stars), Google Code Archive
(jRecruiter\footnote{https://code.google.com/archive/p/jrecruiter/} ), and DARPA
STAC
project \cite{stac}
(TextCrunchr, Braidit, WithMi, Calculator, Battleboats,
Image\_processor, Smartmail, Powerbroker, and Snapbuddy).
To set up the experiments, we created drivers invoking callback methods in
patterns that imitate standard usage.
To support the Object-Oriented feature (which is orthogonal to the problem and approach in the paper), we not only annotate collection-typed local variables, but
also annotate collection-typed object fields that are reachable from local variables.
Then we gave bounds to
each method as tight as possible and used Microsoft Z3 to verify the bounds.

\begin{table}[t]
	\center
\begin{tabular}{l r r r r r r r}
	Benchmarks &
 {Lines} &
 ~~~~~~~~~~~~Verified/ &
  ~~~~Verified/ &
{Call-} &
{Sum- } &
{Analysis} \\
  &
{of Code} &
Unverified  &
Unverified &
 backs &
{maries} &
{time (s)} \\
&
&
methods &
collections &
&
& \\
\hline
TextCrunchr & 2,150 & (13+190)/5 & 23/5  & 4     & 9     & 14.7 \\
Braidit & 20,835 & (8+2114)/0 & 50/0  & 8     & 8     & 84.2 \\
jforum3 & 22,813 & (35+1675)/8 & 54/10 & 24    & 27    & 69.8 \\
jRecruiter & 13,936 & (29+933)/5 & 40/4  & 10    & 7     & 45.1 \\
SpringPetClinic & 1,429 & (6+98)/0 & 11/1  & 9     & 12    & 15.8 \\
WithMi & 24,927 & (30+2515)/5 & 35/2  & 5     & 4     & 82.0 \\
Calculator & 5,378 & (20+316)/2 & 25/6  & 5     & 3     & 18.2 \\
Battleboats & 21,525 & (8+2171)/6 & 12/2  & 5     & 2     & 75.6 \\
Image\_processor & 1,365 & (4+110)/0 & 5/0   & 0     & 0     & 7.8 \\
Smartmail & 1,977 & (7+137)/4 & 11/3  & 0     & 0     & 10.9 \\
Powerbroker & 29,374 & (22+3015)/8 & 27/3  & 3     & 8     & 91.6 \\
Snapbuddy & 34,797 & (57+2940)/8 & 88/9  & 5     & 2     & 107.0 \\
\hline
Total & 180,506 & (239+16214)/51 & 381/45 & 78    & 82    & 622.6
\end{tabular}%
  \vspace*{0.5em}
	\caption{\scriptsize
		Benchmark results.
		``Lines of code'' counts the total lines of code in projects.
		``Verified methods'' gives number of verified methods and unverified methods.
		Number of verified methods is split into \emph{non-vacuously} and \emph{vacuously} verified, where being vacuously verified means not declaring any local collection variables.
		``Verified/Unverified collections'' gives the number of collection variables that are verified versus unverified.
		``Callbacks'' gives the number of invoked callbacks in drivers.
		``Method summaries'' gives the number of method summaries supporting verifying collection variables that are inter-procedurally mutated.
		``Analysis time'' indicates the speed of our analysis on each benchmark.
		Experiments were conducted on a 4-core 2.9 GHz Intel Core i7 MacBook with 16GB of RAM running OS X 10.13.6.}
	\label{tab:experimen-data}
	\vspace{-3em}
\end{table}

\subsection{Research questions}
\label{subsec:research-question}
We evaluated our technique by answering the following research questions
\begin{enumerate}
	\item[RQ1.] \textbf{Bound verification.} How effective is \rcfaname? That is, what percentage of methods and collection variables were verified w.r.t. their specifications.
	\item[RQ2.] \textbf{Analysis speed.} How fast/scalable is our verification
	technique?
\end{enumerate}

\subsubsection{RQ1: Bound verification.} \label{subsubsec:rq1} We verified 16453
methods in total, 239 of which are non-vacuously verified (who declares at least
one collection variable) and the rest are vacuously verified (who declares no
collection variable).
If not considering vacuously verified methods, then we verified 239 out of 290 (82.4\%) methods.
In order to verify method boundedness, we also wrote and verified
global invariants on 381 collection variables out of a total of 426 (89.4\%), as
well as provided 82 method summaries. We invoked 78 callbacks from drivers.
We believe this result demonstrates that our technique is effective at verifying method and variable specifications.
Our technique works very well when there is no statement reading a list-typed element from a collection, which if it happens, constitutes the vast majority of the causes of the 51 unverified methods and 45 unverified collections, because to ensure soundness we had to enforce no constraint on these list-typed variables read from collections, leading to unboundedness.
We currently do not support this feature in the type system and we will leave it for future work.
Note that in the table we did not include unverified methods and variables caused by orthogonal problems such as Java features (e.g. dynamic dispatch) discussed in Section \ref{subsec:limitation-discussion}.

We attribute the effectiveness of \rcfaname\;to the scalable type checking
approach and our AST counter-base approach.
Also note that without \rcfaname, it would have not been possible to \textbf{flow-insensitively} verify the desired properties.
The detailed results from each
benchmark are in Table~\ref{tab:experimen-data}.

\paragraph{Alias.} Note that the operational semantics defined in Figure
\ref{fig:operational-semantic} does not support aliasing among collection-typed
variables. This is because aliasing is orthogonal to the problem and approach in
the paper. To demonstrate that this is indeed the case, we extend our framework
with aliasing, which we present in the Appendix \ref{sec:alias}. The implementation uses the framework from the appendix.

\paragraph{Case studies.} We present an interesting loop that we found and
simplified from the jforum3 benchmark. In the while loop at line 4-17, line 5
reads in a String with \code{readLine} and line 14 adds an element to list
\code{comments}. Although the while loop may not terminate, the inductive
invariant $\len{comments}-\iter{reader}=c5-c14$ is preserved before and every execution
step. Here we consider variable \code{reader} as an iterator, respecting the
semantics of the \code{readLine} API.

\begin{lstlisting}[language=java]
@Inv("len(comments)-idx(reader)=c5-c14") List<String> comments = new ArrayList<>();
// ...
String s;
while (true) {
c5: s = reader.readLine();
	if (s != null) {
		s = s.trim();
	}
	if (s == null || s.length() < 1) {
		continue;
	}
	if (s.charAt(0) == '#') { // comment
	if (collectComments && s.length() > 1) {
		c14: comments.add(s.substring(1));
	}
	continue;
}
\end{lstlisting}

We also discovered a Denial-of-Service bug from benchmark TextCrunchr that is caused by a collection variable with an excessively large length.
TextCrunchr is a text analysis program that is able to perform some useful analysis (word frequency, word length, etc.), as well as process plain text files and compressed files, which it will uncompress to analyze the contents.
The vulnerability is in the decompressing functionality where it uses a collection variable \code{queue} to store files to be decompressed.
Our tool \toolname\;correctly did not verify the boundedness of variable \code{queue} and we believe this leads to TextCrunchr's being vulnerable to a Zip bomb attack, because variable \code{queue} may store an exponential number of files that is caused by a carefully crafted zip file, which contains other carefully crafted zip files inside, thus leading to an exponential number of files to be stored in variable \code{queue} and to be decompressed.

\subsubsection{RQ2: Analysis speed.}
\label{subsubsec: rq2}

On average, it takes 51.9 second to analyze a 15k lines of code benchmark
program (including vacuously verified methods) with \toolname. The detailed results from each benchmark are in
Table~\ref{tab:experimen-data}.  Given the lines of code of our benchmarks, we
believe this result indicates that the speed of our analysis benefits from being
type-based and flow-insensitive, exhibiting the potential of scaling to even
larger programs.

\subsection{Limitations, future work and discussion}
\label{subsec:limitation-discussion}

In experiments, we encountered collection variables that could not be annotated, leaving \toolname\;unable to verify boundedness of methods that declare them.
We next categorize the reasons and discuss future work for improvements:

\begin{itemize}
	\item To ensure soundness, we enforce no constraint on a collection-typed
	variable's length (i.e. allow it to be infinitely long), when it is the result
	of reading a list-typed element from a collection. The reason is that the type
	system does not yet support annotating lengths of inner lists. This extension
	of our type system is left for future work.
	\item Not discovering the right global invariants. In the future we plan to
	automate the invariant discovery process with abstract interpretation, which
	will hopefully help uncover more invariants.
	\item Imprecision and ``soundiness'' caused by Java features such as aliasing,
	dynamic dispatch, inner class, class inheritance and multi-threading.  We
	regard these as orthogonal problems to our problem statement and we could
	extend our type system to support them.
\end{itemize}

\paragraph{Integration with building tools.} To evaluate how user-friendly
\toolname\;is for a developer, we also evaluated how \toolname\;integrates with
open source repositories (i.e. jforum3, jRecruiter and SpringPetClinic) that use
popular building tools (e.g. Maven). We discovered that the configuration is
reasonably easy: Developers only need to add several Maven dependencies
(including \toolname, Checker framework, Scala library, Scala SMT-LIB and
Microsoft Z3's Java bindings) into \code{pom.xml} and specify \toolname\;as an
additional annotation processor. After that, a developer could immediately start
using our tool!

\paragraph{Annotation workflow.} The typical annotation workflow of a user is to
first configure \toolname\;as an annotation processor, and then compile the
target code/project without any annotations. Note that errors and warnings are
expected if \toolname\;cannot prove boundedness of a procedure, which is
intrinsically caused by insufficient annotations (i.e. type refinements). In the
end, a user will fix the errors and warnings by annotating collection variables.
In the case of a method returning a locally allocated collection variable, we inlined the method into its caller to ensure soundness.
Additionally, to perform inter-procedural analysis, we introduce method summaries to
describe changes in lengths of collection-typed variables caused by method
invocation.
Method summaries are expressed in the refinement language defined in Figure
\ref{fig:types-and-refinements}, together with variables in their primed version,
which denotes the values after method invocation. 
Method summaries are automatically applied when type checking a method
invocation statement.
The annotation burden for method summaries was light (6.8 methods on average) in the experiments.


\vspace{-1em}
\section{Related works}
\label{sec:related-works}

\noindent {\bf Type systems for resource analysis.}
Type-based approaches have been proposed for resource analysis~
\cite{chin2001calculating,vasconcelos2003inferring,vasconcelos2008space}. These
works verify size relations between input and output list variables as a
function specification. Additionally, there is a line of works that combines a
type-based approach with the idea of amortized analysis
\cite{hoffmann2011multivariate,hoffmann2017towards} to analyze resource usage.
These approaches are not able to analyze programs with mutation and it is also
unclear how to adapt them into a setting with mutation. The reason is the need
for the analysis to be flow-sensitive in the presence of mutation, because
mutation causes program variables' sizes to change. In contrast, we emphasize
that it is our novelty to introduce \rcfaname, making it possible to write
flow-insensitive types in the presence of mutation, and thus enjoy the benefits
of a type-based approach --- being compositional and scalable. We put back the
limited flow-sensitive information (in the form of counter axioms) only after
the type checking phase.

\noindent {\bf Resource analysis by loop-bound analysis.}
Bound analysis techniques
~\cite{carbonneaux2015compositional,giesl2014proving,gulwani2009control,gulwani2010reachability,sinn2014simple,zuleger2011bound}
emphasize that time bounds (especially loop bounds) are necessary for resource
bound analysis and therefore they focus on obtaining loop bounds.
However, time
boundedness is actually only a sufficient condition for resource boundedness.
In contrast, our approach verifies resource bounds even when time bounds are not available.
The other difference is that, Gulwani et al. and Zuleger et al.'s works
~\cite{gulwani2009control,gulwani2010reachability,zuleger2011bound} generate
invariants at different program locations, as opposed to our approach of using
same invariants at all program locations.
In more detail, Carbonneaux et al. \cite{carbonneaux2015compositional} use a Hoare logic style
inter-procedural reasoning to derive constraints on unknown coefficients of loop bounds,
who are in the form of pre-defined templates consisting of multivariate intervals.
Gulwani et al. \cite{gulwani2009control} introduce
a technique to first transform multi-path loops into loop paths who interleave
in an explicit way, and then generate different invariants at different program
locations. In another work, Gulwani et al. \cite{gulwani2010reachability},
compute the transitive closure of inner loops, which are invariants only hold true
at the beginning of loops. It also utilizes several common loop patterns to
obtain ranking functions, which are eventually used to compute loop bounds. Sinn
et al. \cite{sinn2014simple} flatten multi-path loops into sets of mutual
independent loop paths and uses global lexicographic ranking functions to derive
loop bounds.
Similarly, Giesl et al.
\cite{giesl2014proving} use a standard ranking function approach to obtain loop
bounds for its bound analysis, which is a component of its interleaving of size
analysis and bound analysis.
Zuleger et al. \cite{zuleger2011bound} txake size-change abstraction
approach from termination analysis domain into bound analysis. Size-change
abstraction relates values of variables before and after a loop iteration at the
beginning of a loop, which are eventually used to obtain loop bounds.
Additionally, although Gulwani et al. \cite{gulwani2009speed} also adopt a
counter approach by instrumenting loops with counters, the functionalities of
counters are different. In our \rcfaname\;approach, counters enable writing
flow-insensitive global invariants under the scenario of mutation. In contrast,
the functionality of counters in Gulwani et al.'s work \cite{gulwani2009speed}
is to make it explicit if one loop is semantically (instead of syntactically)
nested in another loop: each loop is associated with a counter and this work
encodes loop nest relations as counter dependencies.


\noindent {\bf Resource analysis by cost-recurrence relations.}
A classical approach to cost analysis
~\cite{albert2007cost,albert2007heap,albert2009live,albert2011more,alonso2012limits,flores2014resource}
is to derive a set of cost recurrence relations from the original program, whose
closed-form solutions will serve as an over-approximation of the cost usage. As
pointed out by Alonso et al. \cite{alonso2012limits}, one of the limitations in
recurrence relation approach is that it poorly supports mutation, because of
ignoring the side effects of a callee that may have on its caller. Since
collection APIs typically have side effects, recurrence relation may not be the
best approach to reason about collection variables' lengths.
Additionally, we believe our approach applies to a wider class of programs, because it is more difficult to find closed-form solutions than our approach of checking if a set of constraints implies the desired property.

\noindent {\bf Refinement types.}
Our type system is inspired by Rondon et al. \cite{rondon2008liquid}. The
subsequent work by Rondon et al. \cite{rondon2010low} propose a flow-sensitive
refinement type system to reason about programs with mutation. Similarly,
Coughlin et al.'s work  \cite{coughlin2014fissile} handles mutation via a
flow-sensitive approach, allowing type refinements to temporarily break and then
get re-established later at some other control locations. It adopts
flow-sensitive analysis between control locations who break and re-establish the
invariant, respectively. Compared with Rondon et al. \cite{rondon2010low} and
Coughlin et al. \cite{coughlin2014fissile}, our work is different because it
separates types and refinements from counter relation axioms, where types and
refinements are flow-insensitive but counter relation axioms are flow-sensitive.
The advantage of our approach over Coughlin et al.'s work is that, to verify a
given property, Coughlin et al.'s work is more expensive because it is
sensitive to the distance (in terms of lines of code) between any two relevant
control locations (i.e., where the first location breaks an invariant and the
second location potentially restores the invariant). More specifically, Coughlin
et al.'s work has to perform flow-sensitive and path-sensitive symbolic
execution between any two relevant control locations. In comparison, our
approach is insensitive to the distance between any two relevant control
locations.

\vspace{-1em}

\section{Conclusion}

We proposed a technique that statically verifies the boundedness of total length
of collection variables when given constraint(s) on input(s). Our technique is
able to verify the above property for non-terminating reactive programs. To
ensure scalability, we take a type-based approach and enforce using global
inductive invariants, as opposed to different invariants at different program
locations. To design global invariants for programs supporting mutation, we
introduce AST counters, which track how many times an AST node was executed. We
then add axioms on relations of the counter variables. Experimental results
demonstrate that our technique is scalable, and effective at verifying bounds.

We plan to build on this work in at least the following two directions: (i)
extending from collection lengths to general memory usage, (ii) generalizing the
AST counter technique and applying it in different contexts.

\newpage

\bibliographystyle{splncs04}


\newpage
\appendix

\section{Base type checking}
\label{sec:base-type-checking}
We present base type checking rules in this section, which guarantees that a type-checked program will not get stuck.
\begin{figure}[H]
	\centering
	\begin{mathpar}
		\inferrule* [Lab={\scriptsize BaseT-Add}]
		{
			\tycheck
			{\tyenv}
			{y: \tlist{\basety}}
			\\\\
			\tycheck
			{\tyenv}
			{x: \basety}
		}
		{
			\tycheck
			{\tyenv}
			{\eadd{y}{x}}
		}
		\inferrule* [Lab={\scriptsize BaseT-Remove}]
		{
			\tycheck
			{\tyenv}
			{y: \tlist{\basety}}
		}
		{
			\tycheck
			{\tyenv}
			{\eremove{y}}
		}
		\inferrule* [Lab={\scriptsize BaseT-Iter}]
		{
			\tycheck
			{\tyenv}
			{y: \tlist{\basety}}
			\\\\
			\tycheck
			{\tyenv}
			{z: \titerator{\basety}}
		}
		{
			\tycheck
			{\tyenv}
			{z = \eiterator{y}}
		}
		\\
		\inferrule* [Lab={\scriptsize BaseT-AssignArith}]
		{
			\tycheck
			{\tyenv}
			{x: \kwint}
			\\\\
			\tycheck
			{\tyenv}
			{e_1: \kwint}
			\\
			\tycheck
			{\tyenv}
			{e_2: \kwint}
		}
		{
			\tycheck
			{\tyenv}
			{x = e_1 \oparith e_2}
		}
		\inferrule* [Lab={\scriptsize BaseT-AssignComp}]
		{
			\tycheck
			{\tyenv}
			{x: \kwbool}
			\\\\
			\tycheck
			{\tyenv}
			{e_1: \kwint}
			\\
			\tycheck
			{\tyenv}
			{e_2: \kwint}
		}
		{
			\tycheck
			{\tyenv}
			{x = e_1 \opcomp e_2}
		}
		\\
		\inferrule* [Lab={\scriptsize BaseT-AssignOr}]
		{
			\tycheck
			{\tyenv}
			{x: \kwbool}
			\\\\
			\tycheck
			{\tyenv}
			{e_1: \kwbool}
			\\
			\tycheck
			{\tyenv}
			{e_2: \kwbool}
		}
		{
			\tycheck
			{\tyenv}
			{x = e_1 \lor e_2}
		}
		\inferrule* [Lab={\scriptsize BaseT-AssignNeg}]
		{
			\tycheck
			{\tyenv}
			{x: \kwbool}
			\\
			\tycheck
			{\tyenv}
			{e: \kwbool}
		}
		{
			\tycheck
			{\tyenv}
			{x = \neg e}
		}
		\\
		\inferrule* [Lab={\scriptsize BaseT-AssignInt}]
		{
			\tycheck
			{\tyenv}
			{x: \kwint}
		}
		{
			\tycheck
			{\tyenv}
			{x = n}
		}
		\inferrule* [Lab={\scriptsize BaseT-AssignBool}]
		{
			\tycheck
			{\tyenv}
			{x: \kwbool}
		}
		{
			\tycheck
			{\tyenv}
			{x = b}
		}
		\inferrule* [Lab={\scriptsize BaseT-AssignIntExpr}]
		{
			\tycheck
			{\tyenv}
			{x: \kwint}
			\\
			\tycheck
			{\tyenv}
			{e: \kwint}
		}
		{
			\tycheck
			{\tyenv}
			{x = e}
		}
		\\
		\inferrule* [Lab={\scriptsize BaseT-AssignBoolExpr}]
		{
			\tycheck
			{\tyenv}
			{x: \kwbool}
			\\\\
			\tycheck
			{\tyenv}
			{e: \kwbool}
		}
		{
			\tycheck
			{\tyenv}
			{x = e}
		}
		\inferrule* [Lab={\scriptsize BaseT-AssignIter}]
		{
			\tycheck
			{\tyenv}
			{x: \titerator{\basety}}
			\\\\
			\tycheck
			{\tyenv}
			{e: \titerator{\basety}}
		}
		{
			\tycheck
			{\tyenv}
			{x = e}
		}
		\inferrule* [Lab={\scriptsize BaseT-AssignList}]
		{
			\tycheck
			{\tyenv}
			{x: \tlist{\basety}}
			\\\\
			\tycheck
			{\tyenv}
			{e: \tlist{\basety}}
		}
		{
			\tycheck
			{\tyenv}
			{x = e}
		}
		\\
		\inferrule* [Lab={\scriptsize BaseT-Next}]
		{
			\tycheck
			{\tyenv}
			{x: \basety}
			\\\\
			\tycheck
			{\tyenv}
			{z: \titerator{\basety}}
		}
		{
			\tycheck
			{\tyenv}
			{x = \enext{z}}
		}
		\inferrule* [Lab={\scriptsize BaseT-Decl}]
		{
			\forall \subastnode{\edecl{\ty}{x} = e}{M}.
			\tycheck{\tyenv}{x = e}
			\\\\
			\tycheck{\tyenv}{\stmt}
		}
		{
			\tycheck
			{\tyenv}
			{
				\overline{\edecl{\ty}{u}}\;
				\overline{\edecl{\ty}{x} = e}\;
				\stmt
			}
		}
		\inferrule* [Lab={\scriptsize BaseT-NewList}]
		{
			\tycheck
			{\tyenv}
			{x: \tlist{\basety}}
		}
		{
			\tycheck
			{\tyenv}
			{x = \enewlist{\basety}}
		}
		\\
		\inferrule* [Lab={\scriptsize BaseT-Block}]
		{
			\overline{\stmt} = \stmt_1;\ldots;\stmt_n
			\\\\
			{\tycheck{\tyenv}{\stmt_i}}
			\text{, for all } i \in \{1,\cdots,n \}
		}
		{
			\tycheck{\tyenv}{\eblock{s}}
		}
		\inferrule* [Lab={\scriptsize BaseT-While}]
		{
			\tycheck{\tyenv}{\stmt}
			\\\\
			\tycheck{\tyenv}{e: \kwbool}
		}
		{
			\tycheck{\tyenv}{\ewhile{\expr}{\stmt}}
		}
		\inferrule* [Lab={\scriptsize BaseT-If}]
		{
			\tycheck{\tyenv}{e: \kwbool}
			\\\\
			\tycheck{\tyenv}{\stmt_1}
			\\
			\tycheck{\tyenv}{\stmt_2}
		}
		{
			\tycheck{\tyenv}{\eif{\expr}{\stmt_1}{\stmt_2}}
		}
		\inferrule* [Lab={\scriptsize BaseT-Skip}]
		{
			{}
		}
		{
			\tycheck{\tyenv}{\eskip}
		}
	\end{mathpar}
	\caption{Base type checking rules}
	\label{fig:basetype-checking-rules}
\end{figure}

\section{Alias}
\label{sec:alias}

In section, we present operational semantics (Figure \ref{fig:operational-semantic-alias}), refinement semantics (Figure \ref{fig:type-semantic-alias}) and type checking rules (Figure \ref{fig:type-checking-rules-alias}) under the scenario of alias, where we will additionally need store $\store$ that serves as a heap.

\begin{figure}[t]
	\centering
	\begin{tabular}{ L p{0.5cm} L }
		\evalty{\env[x]}{\env,\store}{}{\trefine{\basety}{\kwrefine}} & iff &
		\evalty{\env[x]}{\env,\store}{}{\basety} \text{ and }
		\evalrefine{\rftuple{\kwrefine}{x}}{\env,\store}
		\\
		\evalrefine{b}{\env,\store} & iff &
		b \equiv \kwtrue
		\\
		\evalrefine{\rftuple{\sfxbool{y}}{x}}{\env,\store} & iff &     \evalrefine{\env[\sfxbool{y}]}{\env,\store} 
		\\
		\evalrefine{\rftuple{\neg \kwrefine}{x}}{\env,\store} & iff &
		(\evalrefine{\rftuple{\kwrefine}{x}}{\env,\store}) \not = \kwtrue
		\\
		\evalrefine{\rftuple{\kwrefine_1\lor\kwrefine_2}{x}}{\env,\store} & iff &
		\evalrefine{\rftuple{\kwrefine_1}{x}}{\env,\store} \text{ or } \evalrefine{\rftuple{\kwrefine_2}{x}}{\env,\store}
		\\
		\evalrefine{\rftuple{\isiterof{}{y}}{x}}{\env,\store} & iff &
		\text{ for some $i \geq 0$ we have }\env[x] = \iterval{i}{a}
		\text{ and } \store[y] = a
		\\
		\evalrefine{\rftuple{\rfexpr_1\opcomp\rfexpr_2}{x}}{\env,\store} & iff &
		\evalrfexpr{\rftuple{\rfexpr_1}{x}}{\env,\store} \bowtie \evalrfexpr{\rftuple{\rfexpr_2}{x}}{\env,\store}
		\\
		\evalrfexpr{\rftuple{\rfexpr}{x}}{\env,\store} & = &
		v
		\text{, where }
		e = \unintsubst{\rftuple{\rfexpr}{x}}{\env,\store} \land
		\evalexprn{\env}{e}{v}
		\\
		\unintsubst{\rftuple{\rfexpr}{x}}{\env,\store} & = &
		(\rfexpr[x/\kwself])[n_i/\len{\sfxlist{x^i}},k_j/\iter{\sfxiter{y^j}}]_{
			\text{ $\forall \sfxlist{x^i}, \sfxiter{y^j}$}}
		\\ & &
		\text{ where } \env[\sfxlist{x^i}] = a \land \store[a] = \listval{v_1, \ldots, v_{n_i}}
		\\\ & &
		\text{ and } \env[\sfxiter{y^j}] = \iterval{k_j}{*}
	\end{tabular}
	\caption{Refinement semantics. (Alias)}
	\label{fig:type-semantic-alias}
\end{figure}

\begin{figure}[t]
	\scriptsize
	\centering
\subfloat[Environment, Store, and Values] {
		\begin{tabular}{ l L L L }
			Environment & \env & \bnfdef &
			\emp
			\bnfalt \mapext{\env}{x}{v}
			\bnfalt \mapexthook{\env}{\counter}{n}
			\\
			Store & \store & \bnfdef &
			\emp
			\bnfalt \mapext{\store}{a}{l}
			\\
			Values & \val & \bnfdef &
			n \in \mathbb{Z}
			\bnfalt b \in \mathbb{B}
			\bnfalt \iterval{n \in \mathbb{N}}{a \in A}
			\bnfalt a \in A
			\\
			List values & l & \bnfdef &
			\listval{v_1, v_2, \ldots, v_n}
		\end{tabular}
}
\\
\subfloat[Operational semantics]{
	\begin{mathpar}
		\inferrule* [Lab={\scriptsize E-Var}]
		{
			{}
		}
		{
			\evalexpr
			{\env}
			{x}
			{\env[x]}
		}
		\inferrule* [Lab={\scriptsize E-ArithL}]
		{
			\evalexpr
			{\env}
			{\expr_1}
			{\expr_1'}
		}
		{
			\evalexpr
			{\env}
			{\expr_1 {\oparith} \expr_2}
			{\expr_1' {\oparith} \expr_2}
		}
		\inferrule* [Lab={\scriptsize E-ArithR}]
		{
			\evalexpr
			{\env}
			{\expr_2}
			{\expr_2'}
		}
		{
			\evalexpr
			{\env}
			{v {\oparith} \expr_2}
			{v {\oparith} \expr_2'}
		}
		\inferrule* [Lab={\scriptsize E-CompL}]
		{
			\evalexpr
			{\env}
			{\expr_1}
			{\expr_1'}
		}
		{
			\evalexpr
			{\env}
			{\expr_1 {\bowtie} \expr_2}
			{\expr_1' {\bowtie} \expr_2}
		}
		\\
		\inferrule* [Lab={\scriptsize E-CompR}]
		{
			\evalexpr
			{\env}
			{\expr_2}
			{\expr_2'}
		}
		{
			\evalexpr
			{\env}
			{v {\bowtie} \expr_2}
			{v {\bowtie} \expr_2'}
		}
		\inferrule* [Lab={\scriptsize E-OrL}]
		{
			\evalexpr
			{\env}
			{\expr_1}
			{\expr_1'}
		}
		{
			\evalexpr
			{\env}
			{\expr_1 {\lor} \expr_2}
			{\expr_1' \text{ or }  \expr_2}
		}
		\inferrule* [Lab={\scriptsize E-OrR}]
		{
			\evalexpr
			{\env}
			{\expr_2}
			{\expr_2'}
		}
		{
			\evalexpr
			{\env}
			{v {\lor} \expr_2}
			{v {\lor} \expr_2'}
		}
		\inferrule* [Lab={\scriptsize E-Neg}]
		{
			\evalexpr
			{\env}
			{\expr}
			{\expr'}
		}
		{
			\evalexpr
			{\env}
			{\neg \expr}
			{\neg \expr'}
		}
		\\
		\inferrule* [Lab={\scriptsize E-Iterator}]
		{
			\evalcounter{\env}{z = \eiterator{y}}{\env'}
			\\
			\env[y] = a
		}
		{
			\eval
			{\env,\store}
			{}
			{z = \eiterator{y}}
			{\mapext{\env'}{z}{\iterval{0}{a}},\store}
			{}
			{\eskip}
		}
		\\
		\inferrule * [Lab={\scriptsize E-Next}]
		{
			\evalcounter{\env}{x = \enext{z}}{\env'}
			\\
			\env[z] = \iterval{i}{a}
			\\\\
			\store[a] = \listval{v_1,\ldots,v_n}
			\\
			i < n-1
		}
		{
			\eval
			{\env,\store}
			{}
			{x = \enext{z}}
			{
				\mapext
				{\mapext{\env'}{z}{\iterval{i+1}{a}}}
				{x}
				{v_{i+1}}
				,\store
			}
			{}
			{\eskip}
		}
		\\
		\inferrule* [Lab={\scriptsize E-Assign}]
		{
			\evalcounter{\env}{x = \expr}{\env'}
			\\
			\evalexprn{\env}{\expr}{v}
		}
		{
			\eval
			{\env,\store}
			{}
			{x = \expr}
			{\mapext{\env'}{x}{v},\store}
			{}
			{\eskip}
		}
		\inferrule* [Lab={\scriptsize E-Remove}]
		{
			\evalcounter{\env}{\eremove{y}}{\env'}
			\\
			\env[y] = a
			\\
			\store[a] = \listappend{\overline{v}}{v}
		}
		{
			\eval
			{\env,\store}
			{}
			{\eremove{y}}
			{\env',
				\mapext
				{\store}
				{a}
				{\listval{\overline{v}}}
			}
			{}
			{\eskip}
		}
		\\
		\inferrule* [Lab={\scriptsize E-Add}]
		{
			\evalcounter{\env}{\eadd{y}{x}}{\env'}
			\\
			\env[y] = a
			\\
			\store[a] = \listval{\overline{v}}
			\\
			\evalexpr{\env}{x}{v}
		}
		{
			\eval
			{\env,\store}
			{}
			{\eadd{y}{x}}
			{\env',
				\mapext
				{\store}
				{a}
				{\listappend{\overline{v}}{v}}
			}
			{}
			{\eskip}
		}
		\\
		\inferrule* [Lab={\scriptsize E-NewList}]
		{
			\evalcounter{\env}{x = \enewlist{\basety}}{\env'}
			\\
			\env[x] = a
		}
		{
			\eval
			{\env,\store}
			{}
			{x = \enewlist{\basety}}
			{\env',\mapext{\store}{a}{\listval{}}}
			{}
			{\eskip}
		}
		\\
		\inferrule * [Lab={\scriptsize E-IfExpr}]
		{
			\evalcounter{\env}{\eif{\expr}{\stmt_1}{\stmt_2}}{\env'}
			\\
			\evalexpr{\env}{\expr}{\expr'}
		}
		{
			\eval
			{\env,\store}
			{}
			{\eif{\expr}{\stmt_1}{\stmt_2}}
			{\env',\store}
			{}
			{\newcnt{\eif{\expr'}{\stmt_1}{\stmt_2}}}
		}
		\\
		\inferrule * [Lab={\scriptsize E-IfTrue}]
		{
			\evalcounter{\env}{\eif{\kwtrue}{\stmt_1}{\stmt_2}}{\env'}
		}
		{
			\eval
			{\env,\store}
			{}
			{\eif{\kwtrue}{\stmt_1}{\stmt_2}}
			{\env',\store}
			{}
			{\stmt_1}
		}
		\\
		\inferrule * [Lab={\scriptsize E-IfFalse}]
		{
			\evalcounter{\env}{\eif{\kwfalse}{\stmt_1}{\stmt_2}}{\env'}
		}
		{
			\eval
			{\env,\store}
			{}
			{\eif{\kwfalse}{\stmt_1}{\stmt_2}}
			{\env',\store}
			{}
			{\stmt_2}
		}
		\\
		\inferrule * [Lab={\scriptsize E-While}]
		{
			\evalcounter{\env}{\ewhile{\expr}{\stmt}}{\env'}
		}
		{
			\eval
			{\env,\store}
			{}
			{\ewhile{\expr}{\stmt}}
			{\env',\store}
			{}
			{
				\newcnt
				{
					\eif{\expr}
					{
						\stmt;\newcnt{\ewhile{\expr}{\stmt}}
					}
					{\eskip}
				}
			}
		}
		\\
		\inferrule * [Lab={\scriptsize E-Block}]
		{
			\evalcounter{\env}{\eblock{\stmt}}{\env'}
			\\
			\overline{\stmt} = \stmt_1;\stmt_2;\ldots;\stmt_n
			\\\\
			\eval{\env',\store}{}{\stmt_1}{\env'',\store'}{}{\stmt_1'}
			\\
			\overline{\stmt'} = \newcnt{\stmt_1'};\stmt_2;\ldots;\stmt_n
		}
		{
			\eval
			{\env,\store}
			{}
			{\eblock{\stmt}}
			{\env'',\store'}
			{'}
			{\newcnt{\eblock{\stmt'}}}
		}
		\inferrule * [Lab={\scriptsize E-BlockSkip}]
		{
			\overline{\stmt} = \eskip;\stmt_2;\ldots;\stmt_n
			\\
			\overline{\stmt'} = \stmt_2;\ldots;\stmt_n
		}
		{
			\eval
			{\env,\store}
			{}
			{\eblock{\stmt}}
			{\env,\store}
			{}
			{\newcnt{\eblock{\stmt'}}}
		}
		\\
		\inferrule * [Lab={\scriptsize E-Counter}]
		{
			\getcnt{\stmt}{M} = \counter
		}
		{
			\evalcounter
			{\env}
			{\stmt}
			{
				\mapexthook
				{\env}
				{\counter}
				{\env[\counter]+1}
			}
		}
		\inferrule * [Lab={\scriptsize E-Counter-Aux}]
		{
			\getcnt{\stmt}{M} = \bot
		}
		{
			\evalcounter
			{\env}
			{\stmt}
			{\env}
		}
	\end{mathpar}
}
	\caption{Environment, store, values and small-step operational semantics. (Alias)}
	\label{fig:operational-semantic-alias}
\end{figure}

\paragraph{Alias assumption.}
Before introducing type checking rules, we need the below assumption to hold for any reachable state of executing method $M$ to ensure \textbf{soundness}: During execution, any two list-typed variables should either alias with each other from beginning to end, or never alias.
If this assumption
does not hold true for a method, then we consider this method could not be
verified as bounded. In implementation, to obtain alias information between any two
collection-typed variables, we checked if their types are the same, taking
advantage of the fact that generic types are invariant with respect to the
parameterized types in Java, unless bounds are involved. We do not yet consider
type bounds in implementation. Of course, we could obtain more precise alias
information by running an up-front points-to analysis (e.g. WALA \cite{wala}),
but we leave this for future work.


Additionally, we define a helper function $\getalias{x}$ that takes as input a list-typed variable $x$ and returns a set of list-typed variables who \textbf{must} alias variable $x$ at any time during execution.

\begin{figure}[t]
	\centering
	\begin{mathpar}
		\inferrule* [Lab={\scriptsize T-Add}]
		{
			\getcnt{\eadd{y}{x}}{M} = \counter
			\\
			\forall y_i, y_j \in \getalias{y}. \len{y_i} = \len{y_j}
			\\
			\forall (w: \ty_w) \in \tyenv.
			\subtyof
			{(\substitute{\ty_w}{[w/\kwself]})}
			{
				\substitute
				{(\substitute{\ty_w}{[w/\kwself]})}
				{
					[{\len{y'}+1/\len{y'}}^{y' \in \getalias{y}},(\counter+1)/\counter]
				}
			}
		}
		{
			\tycheck
			{\tyenv}
			{\eadd{y}{x}}
		}
		\\
		\inferrule* [Lab={\scriptsize T-Remove}]
		{
			\getcnt{\eremove{y}}{M} = \counter
			\\
			\forall y_i, y_j \in \getalias{y}. \len{y_i} = \len{y_j}
			\\
			\forall (w: \ty_w) \in \tyenv.
			\subtyof
			{(\substitute{\ty_w}{[w/\kwself]})}
			{
				\substitute
				{(\substitute{\ty_w}{[w/\kwself]})}
				{
					[{\len{y'}-1/\len{y'}}^{y' \in \getalias{y}}, (\counter+1)/\counter]
				}
			}
		}
		{
			\tycheck
			{\tyenv}
			{\eremove{y}}
		}
		\\
		\inferrule* [Lab={\scriptsize T-AssignIter}]
		{
			\getcnt{z = \eiterator{y}}{M} = \counter
			\\
			\forall y_i, y_j \in \getalias{y}. \len{y_i} = \len{y_j}
			\\
			\forall (w: \ty_w) \in \tyenv.
			\subtyof
			{(\substitute{\ty_w}{[w/\kwself]})}
			{
				\substitute
				{(\substitute{\ty_w}{[w/\kwself]})}
				{[0/\iter{z},(\counter+1)/\counter]}
			}
			\\
			\tycheck{\tyenv}{z: \ty_z}
			\\
			\subtyof
			{(\substitute{\ty_z}{[z/\kwself]})}
			{
				\substitute
				{(\substitute{\ty_z}{[z/\kwself]})}
				{[0/\iter{z},(\counter+1)/\counter,\isiterof{z}{y}/\isiterof{z}{*}]}
			}
		}
		{
			\tycheck
			{\tyenv}
			{z = \eiterator{y}}
		}
		\\
		\inferrule* [Lab={\scriptsize T-Assign}]
		{
			\getcnt{x = e}{M} = \counter
			\\
			\tycheck{\tyenv}{x: \trefine{\basety}{\kwrefine_x}}
			\\
			\basety \text{ is not a list type}
			\\
			\forall (w: \ty_w) \in \tyenv.
			\subtyof
			{(\substitute{\ty_w}{[w/\kwself]})}
			{
				\substitute
				{(\substitute{\ty_w}{[w/\kwself]})}
				{
					[{e/x},(\counter+1)/\counter]
				}
			}
		}
		{
			\tycheck
			{\tyenv}
			{x = e}
		}
		\\
		\inferrule* [Lab={\scriptsize T-AssignList}]
		{
			\getcnt{x = e}{M} = \counter
			\\
			\tycheck{\tyenv}{x: \trefine{\basety}{\kwrefine_x}}
			\\
			\basety \text{ is a list type}
			\\
			\forall x_i, x_j \in \getalias{x}. \len{x_i} = \len{x_j}
			\\
			\forall (w: \ty_w) \in \tyenv.
			\subtyof
			{(\substitute{\ty_w}{[w/\kwself]})}
			{
				\substitute
				{(\substitute{\ty_w}{[w/\kwself]})}
				{
					[{\len{e}/\len{x'}}^{x' \in \getalias{x}},(\counter+1)/\counter]
				}
			}
		}
		{
			\tycheck
			{\tyenv}
			{x = e}
		}
		\\
		\inferrule* [Lab={\scriptsize T-Next}]
		{
			\getcnt{x = \enext{z}}{M} = \counter
			\\\\
			\forall (w: \ty_w) \in \tyenv.
			\subtyof
			{(\substitute{\ty_w}{[w/\kwself]})}
			{
				\substitute
				{(\substitute{\ty_w}{[w/\kwself]})}
				{
					[\iter{z}+1/\iter{z},(\counter+1)/\counter]
				}
			}
			\\\\
			\forall (w: \trefine{\ty_w}{\kwrefine_w}) \in \tyenv.
			\forall x'\in\getalias{x}.
			\kwrefine_w \Longleftrightarrow \exists x'.\kwrefine_w
		}
		{
			\tycheck
			{\tyenv}
			{x = \enext{z}}
		}
		\\
		\inferrule* [Lab={\scriptsize T-Decl}]
		{
			\evalexpr{\env_\text{init}}{e}{v}
			\\
			\tycheck{\tyenv}{\stmt}
			\\
			\forall \subastnode{\edecl{\ty}{x} = e}{M}. \evalty{v}{\env_\text{init},\store_\text{init}}{}{\ty}
		}
		{
			\tycheck
			{\tyenv}
			{
				\overline{\edecl{\ty}{u}}\;
				\overline{\edecl{\ty}{x} = e}\;
				\stmt
			}
		}
		\\
		\inferrule* [Lab={\scriptsize T-NewList}]
		{
			\getcnt{x = \enewlist{\basety}}{M} = \counter
			\\\\
			\forall (w: \ty_w) \in \tyenv.
			\subtyof
			{(\substitute{\ty_w}{[w/\kwself]})}
			{
				\substitute
				{(\substitute{\ty_w}{[w/\kwself]})}
				{
					[{0/\len{x'}}^{x' \in \getalias{x}},(\counter+1)/\counter]
				}
			}
		}
		{
			\tycheck
			{\tyenv}
			{x = \enewlist{\basety}}
		}
		\\
		\inferrule* [Lab={\scriptsize T-Counter}]
		{
			\stmt \ne \stmt_B
			\\
			\getcnt{\stmt}{M} = \counter
			\\
			\forall (w: \ty_w) \in \tyenv.
			\subtyof
			{\ty_w}
			{
				\ty_w[(\counter+1)/\counter]
			}
		}
		{
			\tycheckcnt
			{\tyenv}
			{\stmt}
		}
		\inferrule* [Lab={\scriptsize T-Skip}]
		{
			{}
		}
		{
			\tycheck{\tyenv}{\eskip}
		}
		\\
		\inferrule* [Lab={\scriptsize T-Block}]
		{
			\overline{\stmt} = \stmt_1;\ldots;\stmt_n
			\\\\
			\tycheckcnt{\tyenv}{\eblock{\stmt}}
			\\
			{\tycheck{\tyenv}{\stmt_i}}
			\text{, for all } i \in \{1,\cdots,n \}
		}
		{
			\tycheck{\tyenv}{\eblock{s}}
		}
		\inferrule* [Lab={\scriptsize T-While}]
		{
			\tycheckcnt{\tyenv}{\ewhile{\expr}{\stmt}}
			\\\\
			\tycheck{\tyenv}{\stmt}
		}
		{
			\tycheck{\tyenv}{\ewhile{\expr}{\stmt}}
		}
		\inferrule* [Lab={\scriptsize T-If}]
		{
			\tycheckcnt{\tyenv}{\eif{\expr}{\stmt_1}{\stmt_2}}
			\\\\
			\tycheck{\tyenv}{\stmt_1}
			\\
			\tycheck{\tyenv}{\stmt_2}
		}
		{
			\tycheck{\tyenv}{\eif{\expr}{\stmt_1}{\stmt_2}}
		}
	\end{mathpar}
	\caption{Type checking rules (Alias)}
	\label{fig:type-checking-rules-alias}
\end{figure}

\section{Proof of Theorem~\ref{theorem:refinement-preservation}}
\label{sec:soundness-proof}

\begin{proof} (of Theorem~\ref{theorem:refinement-preservation}).
	Given $\wellsize{\tyenv}{\env}$ and
	$\tycheck{\tyenv}{\stmt}$ and
	$\eval{\env}{\store}{\stmt}{\env'}{\store'}{\stmt'}$, we focus on proving
	$\wellsize{\tyenv}{\env'}$, because the validity of $\tycheck{\tyenv}{\stmt'}$
	is directly implied from the premises in Figure
	\ref{fig:type-checking-rules}. 
	The goal is to prove for every variable $x_i$ with type $\ty_i$ in
	$\domof{\env}$, we have $\evalty{\env'[x_i]}{\env'}{}{\ty_i[x_i/\kwself]}$. 

	\begin{itemize}
		\item \textbf{Rule \textsc{E-Add}}:
		We need to prove that if $\wellsize{\tyenv}{\env}$ and $\eval{\env}{\store}{\eadd{y}{x}}{\env'}{\store'}{\eskip}$, then
		$\wellsize{\tyenv}{\env'}$.
		From the Rule~\textsc{T-Add}, we have \\
		(Fact 1): $\wellsize{\tyenv}{\env}$ implies
		that  $\wellsize{\tyenv[(\len{y}{+}1)/\len{y}, (\counter{+}1)/\counter]}{\env}$, where we define $\tyenv[(\len{y}{+}1)/\len{y}, (\counter{+}1)/\counter]$ as performing substitution $[(\len{y}{+}1)/\len{y}, (\counter{+}1)/\counter]$ for all types in typing context $\tyenv$.
		\\
		From the Rule~\textsc{E-Add}, we can infer that if
		$\eval{\env}{\store}{\eadd{y}{x}}{\env'}{\store'}{\eskip}$, then
		$\env'(z)=\env(z)$ for variables other than $\counter$ and $y$. Furthermore,
		$\env'[\counter]=\env[\counter]+1$ an
		$\len{\env'[y]}=\len{\env[y]}+1$.
		Based on these properties of $\env'$, we prove by a simple induction on the structure of refinements that \\
		(Fact 2): if $\wellsize{\tyenv[(\len{y}{+}1)/\len{y}, (\counter{+}1)/\counter]}{\env}$ then $\tycheck{\env'}{\tyenv}$. \\
		By chaining Fact 1 and Fact 2, we can conclude the proof.

		\item \textbf{Rule \textsc{E-Remove}}:
		Proof is dual to the proof for Rule \textsc{E-Add}.

		\item \textbf{Rule \textsc{E-Next}}:
		From the Rule~\textsc{T-Next}, we have \\
		(Fact 1): $\wellsize{\tyenv}{\env}$ implies
		that  $\wellsize{\tyenv[(\iter{z}{+}1)/\iter{z}, (\counter{+}1)/\counter]}{\env}$.
		\\
		From the Rule~\textsc{E-Next}, we can infer that if
		$\eval{\env}{\store}{x=\enext{z}}{\env'}{\store'}{\eskip}$, then for every variable $x_i\in\domof{\env}\setminus\{\counter, z, x\}$, we have $\env'[x_i]=\env[x_i]$, $\len{x_i}$, $\isiterof{}{x_i}$ and $\iter{x_i}$ remain unchanged before and after applying Rule \textsc{E-Next}.
		We also have $\env'[\counter]=\env[\counter]+1$, the value of $\iter{z}$ in environment $\env'$ becomes $\iter{z}+1$ where the latter $\iter{z}$ refers to its value in environment $\env$.
		Additionally, $\len{x}/\iter{x}/\isiterof{}{x}$ becomes an arbitrary value.
		Based on these properties of $\env'$, we prove by a simple induction on the structure of refinements that \\
		(Fact 2): if $\wellsize{\tyenv[(\iter{z}{+}1)/\iter{z}, (\counter{+}1)/\counter]}{\env}$ and if $\kwrefine$ does \textbf{not} use variable $x$, then $\tycheck{\env'}{\tyenv}$.
		It is checked by the equivalence between a refinement and its existential quantification of variable $x$ in last premise of Rule~\textsc{T-Next} that if $\kwrefine$ does \textbf{not} use variable $x$.
		\\
		By chaining Fact 1 and Fact 2, we can conclude the proof.

		\item \textbf{Rule \textsc{E-Assign}}:
		From the Rule~\textsc{T-Assign}, we have \\
		(Fact 1): $\wellsize{\tyenv}{\env}$ implies
		that  $\wellsize{\tyenv[e/x, (\counter{+}1)/\counter]}{\env}$.
		\\
		From the Rule~\textsc{E-Next}, we can infer that if
		$\tycheck{\tyenv}{x=\enext{z}}$, we have \\
		$\eval{\env}{\store}{x=e}{\env'}{\store'}{\eskip}$, for every able $x_i\in\domof{\env}\setminus\{\counter, x\}$, we have $\env'[x_i]=\env[x_i]$, $\len{x_i}$, $\isiterof{}{x_i}$ and $\iter{x_i}$ remain unchanged before and after applying Rule \textsc{E-Assign}.
		We also have $\env'[\counter]=\env[\counter]+1$, $\env'[x]=v$ where $\evalexprn{\env}{e}{v}$.
		Based on these properties of $\env'$, we prove by a simple induction on the structure of refinements that \\
		(Fact 2): if $\wellsize{\tyenv[e/x, (\counter{+}1)/\counter]}{\env}$ then $\tycheck{\env'}{\tyenv}$.
		\\
		By chaining Fact 1 and Fact 2, we can conclude the proof.

		\item \textbf{Rule \textsc{E-Assign}} when right hand side is $\enewlist{\basety}$:
		From the Rule~\textsc{T-AssignNewList}, we have \\
		(Fact 1): $\wellsize{\tyenv}{\env}$ implies
		that  $\wellsize{\tyenv[0/\len{x}, (\counter{+}1)/\counter]}{\env}$.
		\\
		From the Rule~\textsc{E-Assign}, we can infer that if
		$\eval{\env}{\store}{x=\enewlist{\basety}}{\env'}{\store'}{\eskip}$, for every able $x_i\in\domof{\env}\setminus\{\counter, x\}$, we have $\env'[x_i]=\env[x_i]$, $\len{x_i}$, $\isiterof{}{x_i}$ and $\iter{x_i}$ remain unchanged before and after applying Rule \textsc{E-Assign}.
		We also have $\env'[\counter]=\env[\counter]+1$ and $\len{x}$ becomes 0.
		Based on these properties of $\env'$, we prove by a simple induction on the structure of refinements that \\
		(Fact 2): if $\wellsize{\tyenv[0/\len{x}, (\counter{+}1)/\counter]}{\env}$ then $\tycheck{\env'}{\tyenv}$.
		\\
		By chaining Fact 1 and Fact 2, we can conclude the proof.

		\item \textbf{Rule \textsc{E-Assign}} when right hand side is a list-typed variable:
		From the Rule~\textsc{T-AssignList}, we have \\
		(Fact 1): $\wellsize{\tyenv}{\env}$ implies
		that  $\wellsize{\tyenv[\len{e}/\len{x}, (\counter{+}1)/\counter]}{\env}$.
		\\
		From the Rule~\textsc{E-Assign}, we can infer that if
		$\eval{\env}{\store}{x=e}{\env'}{\store'}{\eskip}$, for every able $x_i\in\domof{\env}\setminus\{\counter, x\}$, we have $\env'[x_i]=\env[x_i]$, $\len{x_i}$, $\isiterof{}{x_i}$ and $\iter{x_i}$ remain unchanged before and after applying Rule \textsc{E-Assign}.
		We also have $\env'[\counter]=\env[\counter]+1$ and $\len{x}$ becomes $\len{e}$.
		Based on these properties of $\env'$, we prove by a simple induction on the structure of refinements that \\
		(Fact 2): if $\wellsize{\tyenv[\len{e}/\len{x}, (\counter{+}1)/\counter]}{\env}$ then $\tycheck{\env'}{\tyenv}$.
		\\
		By chaining Fact 1 and Fact 2, we can conclude the proof.

		\item \textbf{Rule \textsc{E-Assign}} when right hand side is $\eiterator{y}$:
		From the Rule~\textsc{T-AssignIter}, we have \\
		(Fact 1): $\wellsize{\tyenv}{\env}$ implies
		that $\wellsize{\tyenv[0/\iter{z}, (\counter{+}1)/\counter]}{\env}$.
		\\
		From the Rule~\textsc{E-Assign}, we can infer that if
		$\eval{\env}{\store}{x=\eiterator{y}}{\env'}{\store'}{\eskip}$, for every variable $x_i\in\domof{\env}\setminus\{\counter, x\}$, we have $\env'[x_i]=\env[x_i]$, $\len{x_i}$, $\isiterof{}{x_i}$ and $\iter{x_i}$ remain unchanged before and after applying Rule \textsc{E-Assign}.
		We also have $\env'[\counter]=\env[\counter]+1$, $\iter{z}$ becomes 0, and any type refinement of form $\isiterof{}{*}$ for variable $z$ becomes $\isiterof{}{y}$.
		Based on these properties of $\env'$, we prove by a simple induction on the structure of refinements that \\
		(Fact 2): if $\wellsize{\tyenv[0/\iter{z}, (\counter{+}1)/\counter]}{\env}$ then $\tycheck{\env'}{\tyenv}$.
		\\
		By chaining Fact 1 and Fact 2, we can conclude the proof.

		\item \textbf{Rule \textsc{E-IfExpr}}: Proof is trivial because only the counter associated with statement $\eif{\expr}{\stmt_1}{\stmt_2}$ is incremented, which is checked by the premise that uses Rule \textsc{T-Counter}.

		\item \textbf{Rule \textsc{E-IfTrue}} and \textbf{Rule \textsc{E-IfFalse}}: Proof is trivial because every type refinement remains unchanged.

		\item \textbf{Rule \textsc{E-While}}: Proof is trivial because only the counter associated with statement $\ewhile{\expr}{\stmt}$ is incremented, which is checked by the premise that uses Rule \textsc{T-Counter}.

		\item \textbf{Rule \textsc{E-Block}}: Proof is trivial because only the counter associated with statement $\eblock{\stmt}$ is incremented, which is checked by the premise that uses Rule \textsc{T-Counter}.

		\item \textbf{Rule \textsc{E-BlockSkip}}: Proof is trivial because every type refinement remains unchanged.

		\item \textbf{Rule \textsc{E-Var}, \textsc{E-ArithL}, \textsc{E-ArithR}, \textsc{E-CompL}, \textsc{E-CompR}, \textsc{E-OrL}, \textsc{E-OrR}, \textsc{E-Neg}, \textsc{E-Iterator}, \textsc{E-NewList}}: Proof is trivial because every type refinement remains unchanged.
	\end{itemize}
\end{proof}

\end{document}